\documentclass[%
 preprint,
 amsmath,amssymb,
 aps, physrev,
]{revtex4-2}

\usepackage[table]{xcolor}
\usepackage{longtable}
\usepackage{graphicx} % figures
\usepackage{amssymb,amsmath,amsfonts,amsthm} % Standard AMS math symbols
\usepackage{color} % text colors
\usepackage{siunitx} % use properly formatted SI units

\usepackage{url} % create hyperlinks
\usepackage{enumerate} % remove excessive list spacing
\usepackage{tabularx}
\newtheorem{remark}{Remark}

\begin{document}

\preprint{APS/123-QED}

\renewcommand*{\arraystretch}{.4} % compress tables.

\title{A Mathematical Model of Hematopoiesis during Systemic Infection}

\author{Bryan Bailey-Feliciano}
\affiliation{Department of Mathematics, University of Florida, Gainesville, FL 32611}
\author{George W. VanVeckhoven}
\affiliation{Department of Mathematics, University of Florida, Gainesville, FL 32611}
\author{Yongchen Tai}
\affiliation{Department of Mechanical and Aerospace Engineering, University of Florida, Gainesville, FL 32611}
\author{Jing Pan\footnote{Corresponding author jingpan@ufl.edu}}
\affiliation{Department of Mechanical and Aerospace Engineering, University of Florida, Gainesville, FL 32611}
\author{Youngmin Park\footnote{Corresponding author park.y@ufl.edu}}
\affiliation{Department of Mathematics, University of Florida, Gainesville, FL 32611}

%\maketitle

\date{\today}% It is always \today, today,
             %  but any date may be explicitly specified

\begin{abstract}Chronic critical illness (CCI) is a disease state in which, following an initial insult, a patient neither recovers nor dies but instead remains in a state of critical illness. CCI is characterized by prolonged organ dysfunction, weight loss, and persistent increased vulnerability to infection. Recent data has shown that patients with CCI generally exhibit persistent immune dysfunction, characterized by prolonged elevation of specific pro-inflammatory cytokines. In this paper, we introduce a host response model that couples hematopoiesis dynamics with the immune response to infection. Specifically, we incorporate the reactions between pro-inflammatory and anti-inflammatory signals with specific hematopoietic stem cell compartments with a reduced model of acute inflammation. We found that a maladaptive hematopoietic response to pathogenic insult is able to qualitatively reproduce similar behavior to that seen in CCI patients, namely the presence of a persistent, elevated level of pro-inflammatory cytokines. This suggests that maladaptive hematopoietic responses \textit{in vivo} may play a role in the development of CCI.
\end{abstract}

\maketitle

\section{Introduction}

 Sepsis is one of the most deadly and expensive diseases to manage in intensive care units (ICUs) in the United States \cite{afsharCCM2019}. Despite the decrease of in-patient mortality over the past decades, current epidemiology shows many sepsis survivors often develop chronic critical illness (CCI), resulting in poor patient outcomes \cite{fennerFM2020}. Understanding the molecular and cellular mechanisms leading to CCI is an important step towards improving patient management.

 In healthy individuals, the immune system is in homeostasis. WBC production (leukopoiesis) and blood-cell production (hematopoiesis) are at basal levels. However, external insults (e.g. systemic infection, trauma) change the body's demand for immune effector cells, which is signaled through physiological processes such as inflammation. Such physiological changes, in turn, alter hematopoietic homeostasis to prioritize the production of WBCs at the expense of red blood cells (erythrocytes) to meet the increased immunologic demand. For patients with an impaired immune system (e.g. those with comorbidities), the deviation from immune and hematopoietic homeostasis can cause both acute \cite{angusNEJM2013} and chronic \cite{carmichaelAS2022} physiologic disorders, which eventually result in poor patient outcomes. In the biological literature, it has been proposed that hematopoiesis insufficiency and emergency myelopoiesis generate immature myeloid-derived suppressor cells (MDSC) as the main drivers of physiological disruptions (pathophysiologic processes) observed in CCI patients \cite{rinconJLB2022}.  

 Failure to identify abnormal immune system function due to dysfunctional hematopoiesis is a major challenge for patient management in the clinic \cite{rinconJLB2022, carmichaelAS2022, torresARP2022}. While clinical evidence has established that CCI patients have elevated levels of circulating inflammatory signaling molecules \cite{stortzS2018, hawkins_chronic_2018}, it is still unclear what biomarkers are most specific to disease progression and how often measurements should be taken for accurate prognosis and precise intervention. There is a critical need to elucidate the dynamical changes in hematopoiesis during sepsis. To this end, various models have been constructed to explore aspects of the acute immune response, which we will briefly summarize (for more detailed information on the sepsis modeling literature, we refer the reader to \cite{reynolds2008mathematical,vodovotz2020silico,schuurman2023embracing}). In particular, we focus on mechanistic models. For data-driven or model-free methods, see \cite{bara2016model,bara2018toward,ramirez2022data}.
 
 The acute immune response is often modeled using ordinary differential equations to describe key quantities involved in sepsis. Typical quantities include pathogen load, pro- and anti-inflammatory signaling molecules, and tissue damage. The detail to which these quantities are modeled and how they interact depends on the particular application and available data. For instance, \cite{kumar2004dynamics} and \cite{barber2021predicting} aim to reproduce data in animal models of sepsis, and successfully reproduce recovery, aseptic death, or septic death using idealized three- or four-dimensional models of the immune response. These relatively low-dimensional models are useful because they can be straightforward to analyze and understand. However, additional detail is often necessary to better understand mechanisms of the acute immune response, such as whole-body or multi-organ interactions \cite{chow2005acute,mcdaniel2019whole,dobreva2021physiological,relouw2024mathematical}, for the purpose of improving clinical outcomes. Additional details are also useful for highly specific modeling, such as in \cite{presbitero2018supplemented}, where surgeons modeled the detailed effects of alkaline phosphate for use in cardiac surgery, and found a plausible mechanism for how alkaline phosphate may reduce inflammation during surgery. In turn, this model and mechanism can be used to test different administration options for alkaline phosphate during surgery.

 The question of mechanisms is important enough to warrant the study of models without directly calibrating on data. The work by Reynolds et al. \cite{reynolds2006reduced,reynolds2008mathematical} and Day et al. \cite{day2006reduced} highlights this approach, where multiple organ interactions are examined as part of a mechanistic analysis of sepsis. More recent works consider additional mechanisms, such as energy consumption \cite{ramirez2019mathematical}, chemotaxis \cite{giunta2021pattern}, and long-term immunity against specific pathogens (adaptive immunity) \cite{shi2015mathematical,jonas2023modeling}.

 %In \cite{reynolds2008mathematical}, Reynolds et al. constructed and analyzed a set of reduced multi-compartment models of inflammation to better understand mechanisms of multiple organ failure during sepsis. In \cite{relouw2024mathematical}, Relow et al. developed a model to understand the effect of short and long-term exposure to pathogenic insults on the expression of specific immune signaling molecules (such as cytokines), and their impact on heart rate, blood pressure, tissue damage, and temperature regulation. In \cite{dobreva2021physiological}, Dobreva et al. included pain along with cardiovascular and thermal dynamics in their model of acute inflammation. In \cite{almansour2024modelling}, Almansour et al. considered a continuum of macrophage phenotypes from pro- to anti-inflammatory phenotypes to study the extent that macrophage-mediator interactions play in the occurrence of stable chronic inflammatory conditions.

 Although these models provide valuable information on how the inflammatory system interacts with organs, white blood cells (WBCs), and cytokines, they do not consider the hematopoietic process to play a key role in inflammation. In this study, we present a relatively detailed 12-variable mechanistic model of hematopoiesis, modeling the role hematopoietic stem cells (HSCs) play in maintaining cell populations at steady-state and during infection. To achieve this, we bridge models of leukopoiesis, a subset of hematopoiesis focusing specifically on the generation of WBCs from HSCs, an acute infection response model to infection, and an inflammation signaling response. With this model, we seek to gain insight into the relationship between the maladaptive hematopoietic response during acute, systemic infection and the subsequent chronic disease in a patient.

 The paper is organized as follows. We introduce the model in Section \ref{sec:model} and include details of its derivation in Appendix \ref{si:model}. We then show the results of the model in Section \ref{sec:results}, with plots of key behaviors in Section \ref{sec:model_replicates} and a sensitivity analysis in Section \ref{sec:sensitivity}. We conclude with a discussion of our results in Section \ref{sec:discussion}.

\section{Model}\label{sec:model}

The main mechanism that this model aims to qualitatively recreate is leukopoiesis under steady-state health conditions and under infection-driven inflammatory processes. Leukopoiesis is a subset of hematopoiesis in which white blood cells are formed from hematopoietic stem cells in the bone marrow. We model the dynamic behavior between inflammatory signals, leukopoiesis, and infection. The model consists of 4 types of compartments (Figure \ref{fig:nodes_hematopoiesis}):
\begin{enumerate}
    \item \textbf{Cells}: The main actors of the system that directly respond to inflammatory signals and can influence their local inflammatory environment through a variety of mechanisms
    \item \textbf{Molecular Mediators}: Secreted by cells or surrounding tissue to mediate communication between cells and sustaining cellular populations
    \item \textbf{Pathogens}: Do not respond to or secrete cytokines, induces a pro-inflammatory response, can kill (i.e. down-regulate) hematopoietic cells
    \item \textbf{Tissue Damage}: An abstract proxy variable that serves as a measurement of overall system health, it affects the maximum potential carrying capacity of certain cell groups
\end{enumerate}

\subsection{Derivation}

We derive the model in sequence, starting with the hematopoietic process, and detail how we incorporate additional features including the acute inflammatory response to infection, tissue damage, nutrients, and myeloid-derived suppressor cell dynamics. Skip to Section \ref{sec:model_equations} for the full model equations simulated in this paper, and Appendix \ref{si:parameters} for the model parameters.

\subsubsection{Modeling the Hematopoietic Process}\label{si:hematopoietic}

 To model the relationship between hematopoiesis and the acute inflammatory response, we focus on a subset of hematopoiesis - leukopoiesis. Leukopoiesis is the process by which hematopoietic stem / progenitor cells, hereby abbreviated HSPCs, divide and up-regulate all classes of white blood cells (WBCs). The process of leukopoiesis begins with HSPCs. HSPCs are a heterogeneous class of stem cells that carry the capacity to self-renew and differentiate into all blood cell lineages, an ability that is unique to them and is incrementally lost as the cell divides and specializes \cite{schulz2009hematopoietic} \cite{laurenti2018haematopoietic}. These cells are retained in a specialized environment known as the Bone Marrow Niche (hereby abbreviated as BM Niche) and, under steady-state conditions, the majority of these cells are held in a  \textit{$G_0$} phase in which they are not actively dividing. Upon interaction with a number of signals from the local inflammatory environment, HSPCs exit quiescence and enter an active differentiation \textit{$G_1$} phase. 

 We detail the derivation of the equations governing our choice to divide leukopoiesis into 4 stages. Sorted by increasing commitment/maturation they are:
 
\begin{enumerate}
    \item \underline{$H_{ST}$}: Short-term HSPCs, stem cells that have exited quiescence and have enhanced cell-cycling times and limited self renewal; these are functionally distinct from long-term HSPCs in that long-term HSPCs are maintained in quiescence for very long periods of time, maintain extensive self-renewal potential throughout their whole life, and are used to maintain cell populations throughout an entire lifetime whereas short-term HSPCs have diminished self-renewal capacity, are more active and proliferative, and are more involved with up-regulating blood cell populations on the time-scale of weeks to months; because our model is focused on shorter time spans on the order of weeks to months, we focused on ST-HSPCs rather than LT-HSPCs \cite{spangrude1988purification} \cite{li2012pth}
    \item \underline{$MPP$}: Short-term HSPCs that have differentiated into \textit{multipotent progenitor} cells (or MPP for short); these cells have mostly lost their ability to self-renew, have an increased self-cycling rate, are much more reactive to signals from the local inflammatory environment, and serve as the base (or progenitor) for the next step of hematopoiesis
    \item \underline{$S$}: Immature WBCs; these white blood cells have a reduced capacity for cytokine signaling and pathogen removal compared to their mature pro-inflammatory counterparts and serve as the basis from which all of the mature WBC compartments draw from to increase their numbers.
    \item \underline{$Q$ and $U$}: The last stage of maturation; for our reduced hematopoiesis model we simplify WBCs into either a pro-inflammatory phenotype $Q$ or anti-inflammatory phenotype $U$.
\end{enumerate}

\begin{figure}[ht!]
\centering
  \includegraphics[width=\textwidth]{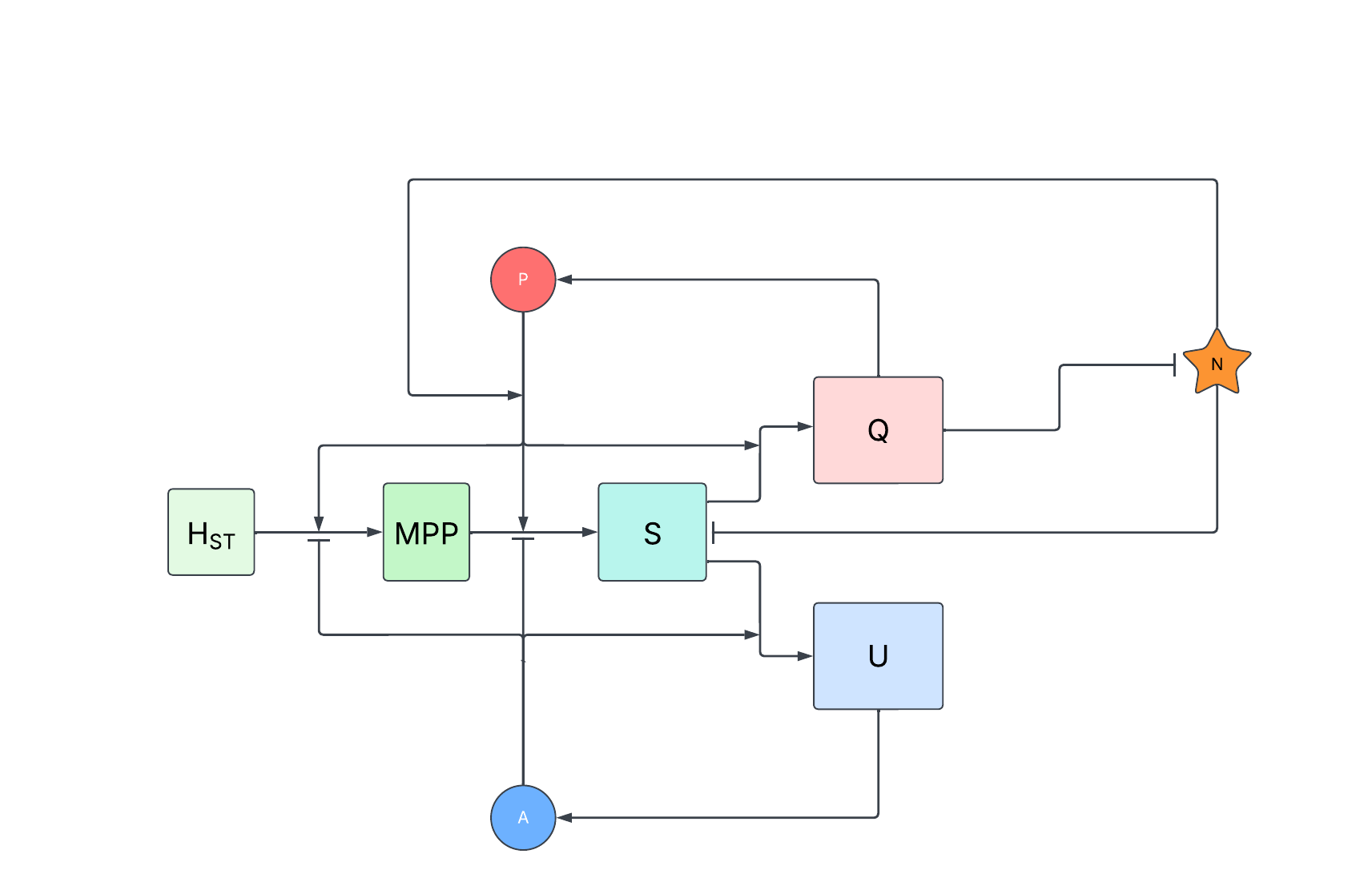}
  \caption{Leukopoiesis divided into 4 discrete stages of increasing differentiation; pro-inflammatory signals induce leukopoiesis while anti-inflammatory signals inhibit it.}\label{fig:nodes_hematopoiesis}
\end{figure}

 The acute inflammatory response is a highly coordinated and complex biological process mediated by molecular signals including chemokines, which direct WBC migration toward sites of infection, and cytokines, a highly diverse group of molecules that facilitate intercellular communication to orchestrate successive phases of the immune response. Because we are not explicitly modeling the migration of WBCs, we largely focus on cytokines and the role they play in coordinating the timing of the immune response. Cytokines, as a group, are very broad and many cytokines are classified as \textit{pleiotropic}, that is they can exert both pro-inflammatory and anti-inflammatory effects depending on the specific context. An example of this is IL-6, which is well known for its pro-inflammatory role in acute and chronic inflammation \cite{gabay2006interleukin}, B and T cell differentiation \cite{maeda20106}, and its anti-apoptotic effects on T cells \cite{atreya2000blockade} all of which are pro-inflammatory aligned effects, but it has also been shown to promote tissue regeneration in intestinal epithelial cells \cite{scheller2011pro} and to control the extent of acute pro-inflammatory responses \cite{xing19986}. For our model, we simplify this complexity using a description of inflammatory signals composed of a \textbf{pro-inflammatory signal} compartment and an \textbf{anti-inflammatory signal} compartment (denoted by the variables $P$ and $A$ respectively). Where the former enhances the rate of differentiation, the latter inhibits it. This effect is summarized with the ``crowding'' function,
\begin{equation}\label{eq:crowding0}
    I \equiv I(P, A) = \frac{P}{P + A},
\end{equation}
 which will be utilized throughout the model to modulate differentiation rates that are sensitive to an excess of anti-inflammatory cytokines. We call \eqref{eq:crowding0} a ``crowding'' function because the anti-inflammatory signals can crowd out the pro-inflammatory signals.

 \begin{remark}
     Note that we will later alter \eqref{eq:crowding0} to include additional effects from pathogens ($N$) and damage associated molecular patterns (DAMPs), so that \eqref{eq:crowding0} becomes, 
    \begin{equation}\label{eq:crowding1}
        I \equiv I(P, A, N, K) = \frac{\text{Amp}_{PN}(N) \cdot \text{Amp}_{PK}(K) \cdot P}{\text{Amp}_{PN}(N) \cdot \text{Amp}_{PK}(K) \cdot P + \text{Amp}_{AK}(K) \cdot A},
    \end{equation}
    where the functions $\text{Amp}_{ij}$ are defined below in \eqref{eq:I_functions}. So, the derivation in this section (\ref{si:hematopoietic}) will proceed using $I$ as shown in \eqref{eq:crowding0}, or $I$ in \eqref{eq:crowding1} assuming $K$ and $N$ are constant.
 \end{remark}

 HSPCs (represented by the state variable $H_{ST}$) and the BM niche are sensitive to various pro-inflammatory signals (e.g. TNF-$\alpha$, IL-1, IL-6, etc.). These signals result in the activation and differentiation of HSPCs, and their exit from the BM niche into the local tissue environment. The rate at which HSPCs differentiate upon binding with pro-inflammatory molecules can be modeled as
\begin{equation}\label{eq:hspc_diff_rate_0}
    \frac{\alpha \cdot P \cdot H_{ST}}{\alpha \cdot P + H_{ST}},
\end{equation}
 where $\frac{1}{\alpha}$ represents the number of pro-inflammatory molecules needed per $H_{ST}$ to induce mobilization (e.g. if 3 $P$ are required to mobilize 1 $H_{ST}$, then $\alpha = 1/3$); in other words, it represents how \textit{sensitive} HSPCs are to pro-inflammatory signals. 
 
 Because a large number of anti-inflammatory signals may inhibit the HSPC differentiation rate, we include the crowding function \eqref{eq:crowding0} into \eqref{eq:hspc_diff_rate_0} to yield the rate
\begin{equation}\label{eq:hspc_diff_rate_1}
    \frac{\alpha \cdot P \cdot I \cdot H_{ST}}{\alpha \cdot P \cdot I + H_{ST}}.
\end{equation}
 With this change, a large number of pro-inflammatory molecules alone is not always sufficient to induce a high differentiation rate in \eqref{eq:hspc_diff_rate_1}; if the anti-inflammatory force significantly out-competes, i.e. outnumbers, the pro-inflammatory force, the rate of differentiation will be inhibited regardless of whether there are enough pro-inflammatory molecules per HSPC to induce differentiation.

 Multipotent progenitor cells (represented by the state variable $MPP$) begin two types of differentiation upon binding with pro-inflammatory signals: symmetric (1 parent HSPC $\rightarrow$ 2 daughter WBCs) or asymmetric (1 parent HSPC $\rightarrow$ 1 daughter HSPC + 1 daughter WBC). The rate of symmetric differentiation is given by,
\begin{equation*}
    \frac{\alpha \cdot P \cdot I \cdot MPP}{\alpha \cdot P \cdot I + MPP} \cdot \frac{I}{I + I_{\text{crit}}},
\end{equation*}
 and the rate of asymmetric differentiation is given by,
\begin{equation*}
    \frac{\alpha \cdot P \cdot I \cdot MPP}{\alpha \cdot P \cdot I + MPP} \cdot \left[1 - \frac{I}{I+ I_{\text{crit}}}\right].
\end{equation*}
 In other words, we model the differentiation rate of multipotent progenitor cells to not only be inhibited by an excess of anti-inflammatory signals, but we allow the crowding function \eqref{eq:crowding0} to determine the proportion of multipotent progenitor cells that become symmetric or asymmetric. So, for instance, if $P\ll A$, then the few multipotent progenitor cells that differentiate will exhibit asymmetric differentiation. On the other hand, if $A \ll P$, then there will be a high rate of multipotent progenitor cells that differentiate, and they will exhibit symmetric differentiation. The parameter $I_\text{crit}$ ($0 < I_\text{crit} \leq 1$) is a free parameter that gives us some control over this symmetric or asymmetric differentiation. To accurately simulate the significantly diminished self-renewal capacity of MPPs, we typically choose $0.5 \leq I_{crit}$.
 
 Immature stable cells ($S$) are sensitive to both pro- and anti-inflammatory signals, and will differentiate into a pro-inflammatory phenotype ($Q$) or an anti-inflammatory phenotype ($U$). The differentiation rate into the pro-inflammatory phenotype is given by,
\begin{equation*}
    \frac{(\tau_QP + \tau_U A) \cdot \Psi S}{\tau_QP + \tau_U A + \Psi S} \cdot \frac{\tau_Q P}{\tau_Q P + \tau_U A},
\end{equation*}
 and the differentiation rate into the anti-inflammatory phenotype is given by,
\begin{equation*}
    \frac{(\tau_QP + \tau_U A) \cdot \Psi S}{\tau_QP + \tau_U A + \Psi S} \cdot \frac{\tau_U A}{\tau_Q P + \tau_U A}.
\end{equation*}
 The parameters $\tau_Q$ and $\tau_U$ control the sensitivity of the immature stable cells $S$ to the pro- and anti-inflammatory signals $P$ and $A$, respectively. The parameter $\Psi$ is a free parameter that gives us some control over the rate of differentiation from $S$ to $Q$ or from $S$ to $U$.

 We thus summarize the equations describing the hematopoietic process (Figure \ref{fig:nodes_hematopoiesis}):
 \begin{equation}\label{eq:hematopoiesis0}
    \begin{split}
        \frac{dH_{ST}}{dt} &= - \frac{\alpha \cdot I \cdot P \cdot H_{ST}}{\alpha \cdot I \cdot P + H_{ST}}, \\
    \frac{dMPP}{dt} &= \frac{\alpha \cdot I \cdot P \cdot H_{ST}}{\alpha \cdot I \cdot P + H_{ST}} - \frac{\alpha \cdot P \cdot I \cdot MPP}{\alpha \cdot P \cdot I + MPP} - d_H MPP,\\
    \frac{dS}{dt} &= \frac{\alpha \cdot P \cdot I \cdot MPP}{\alpha \cdot P \cdot I + MPP} \cdot \left[1 + \frac{I}{I + I_{\text{crit}}}\right] - \frac{(\tau_QP + \tau_U A) \cdot \Psi S}{\tau_QP + \tau_U A + \Psi S} - d_S S,\\
    \frac{dQ}{dt} &= \frac{(\tau_QP + \tau_U A) \cdot \Psi S}{\tau_QP + \tau_U A + \Psi S} \cdot \frac{\tau_Q P}{\tau_Q P + \tau_U A} - d_Q Q,\\
    \frac{dU}{dt} &= \frac{(\tau_QP + \tau_U A) \cdot \Psi S}{\tau_QP + \tau_U A + \Psi S} \cdot \frac{\tau_U A}{\tau_Q P + \tau_U A} - d_U U,\\
    \frac{dP}{dt} &= S_{PH}MPP + S_{PS}S + S_{PQ}Q - d_P P,\\
    \frac{dA}{dt} &= S_{AH}MPP + S_{AS}S + S_{AU}U - d_A A.
    \end{split}
\end{equation}
System \eqref{eq:hematopoiesis0} is our baseline hematopoiesis model, which we now augment with additional dynamics for pathogens, tissue damage, and myeloid-derived suppressor cells. We detail the addition of new variables in sequence below. \textbf{Note that \eqref{eq:hematopoiesis0} is not the final form of our model}. In the process of the derivation in this section, we will alter the growth rates of $Q$ and $U$, alter the differentiation rate of $S$, in ways that depend on new variables for pathogens ($N$), damage associated molecular patterns or DAMPs ($K$), stem cell supporting factors ($SCSF$), essential nutrients ($EN$), and myeloid-derived suppressor cells ($MDSC$).

\subsubsection{The Acute Inflammatory Response to Infection}\label{si:acute_respose}

\begin{figure}[ht!]
\centering
  \includegraphics[width=\textwidth]{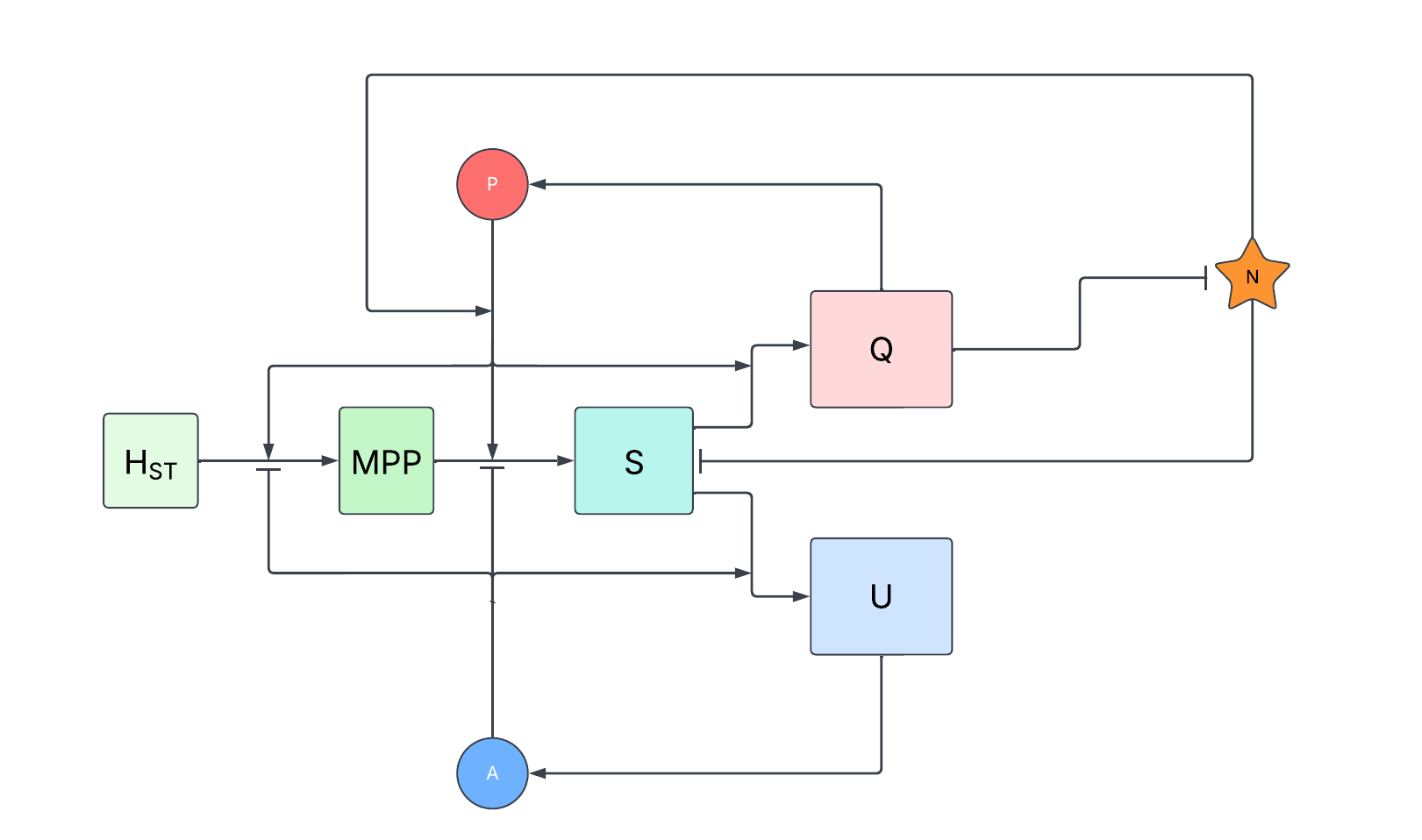}
  \caption{Pathogens (N) amplify the pro-inflammatory immune response.}\label{fig:nodes_acute}
\end{figure}

 We use a logistic growth equation of the form
\begin{equation*}
    \frac{dN}{dt} = g_N N (1 - \frac{N}{N_{\infty}}),
\end{equation*}
 where $g_N$ is the growth coefficient and $N_{\infty}$ is the carrying capacity of pathogens in the system. With the addition of pathogens as a state variable (hereby abbreviated with the state variable \textbf{N}), we can now model the down-regulatory effect pathogens exert on the cells and vice versa. We do this as follows

\begin{equation*}
    \frac{dN}{dt} = g_N N \left[1 - \frac{N}{N_{\infty}}\right] - \frac{(k_{nq}Q + k_{ns}S)N}{N + N_{1/2}}
\end{equation*}

 where $k_{nj}$ represents the kill rate of $N$ per cell $j$. Notice that we multiply $\frac{N}{N + N_{1/2}}$ into the down-regulatory term. This is important as it modifies the rate at which pathogens are eliminated and serves both as a way to ensure that as $N \rightarrow 0$, $(k_{nq}Q + k_{ns}S)(\frac{N}{N + N_{1/2}}) \rightarrow 0$ (preventing $N$ from going negative in our expected cases) as well as a way to model the elusiveness of the pathogen, or how well the pathogen is at evading the immune system with higher values for $N_{1/2}$ correlating to higher evasive capability. 
 
 To model how pathogens can kill i.e. directly down-regulate cells, we modify the term 
\begin{equation*}
     \frac{(\tau_QP + \tau_U A) \cdot \Psi S}{\tau_QP + \tau_U A + \Psi S}
\end{equation*}
 in \eqref{eq:hematopoiesis0} by adding the term $k_{sn} N$ where $k_{sn}$ is the kill rate of $S$ per $N$, thus modifying $S$, $Q$, $U$ in \eqref{eq:hematopoiesis0} into
\begin{align*}
    \frac{dS}{dt} &= \frac{\alpha \cdot P \cdot I \cdot MPP}{\alpha \cdot P \cdot I + MPP} \cdot \left[1 + \frac{I}{I + I_{\text{crit}}}\right]  - \frac{(\tau_QP + \tau_U A + k_{sn} N) \cdot \Psi S}{\tau_QP + \tau_U A + k_{sn} N + \Psi S} - d_S S,\\
    \frac{dQ}{dt} &= \frac{(\tau_QP + \tau_U A + k_{sn} N) \cdot \Psi S}{\tau_QP + \tau_U A + k_{sn} N + \Psi S} \cdot \frac{\tau_Q P}{\tau_Q P + \tau_U A + k_{sn} N} - d_Q Q,\\
    \frac{dU}{dt} &= \frac{(\tau_QP + \tau_U A + k_{sn} N) \cdot \Psi S}{\tau_QP + \tau_U A + k_{sn} N + \Psi S} \cdot \frac{\tau_U A}{\tau_Q P + \tau_U A + k_{sn} N} - d_U U.
\end{align*}

 Furthermore, it is known that many immune cells, like macrophages and HSPCs, express pattern recognition receptors (PRRs) and toll-like receptors (TLRs) on their surface in order to detect and respond to pathogens. Once a pathogen associated molecular pattern (PAMP) binds to one of these receptors, a number of different effects can occur. In our reduced model, we simply assume that they all contribute to a pro-inflammatory immune response. We model this by defining the function $\text{Amp}_{PN}(N)$ in the crowding function \eqref{eq:crowding1} as
\begin{equation*}
    \text{Amp}_{PN}(N) = \frac{N^k}{N^k + \theta_N} + 0.25,
\end{equation*}
 where $k$ is a hill-type coefficient and $\theta_N$ represents the concentration at which this reaches half of its maximum velocity. The $0.25$ is largely arbitrary -- the value for this term modifies the strength of the pro-inflammatory force when pathogens are absent. Hence a value of $0.25$ implies that in the absence of pathogens stimulating the PPR and TLR receptors, the pro-inflammatory force is dampened down to a quarter of its strength.

\subsubsection{Tissue Damage, Cellular Nutrients, \& Carrying Capacities}\label{si:damage}

\begin{figure}[ht!]
\centering
  \includegraphics[height=0.7\textwidth]{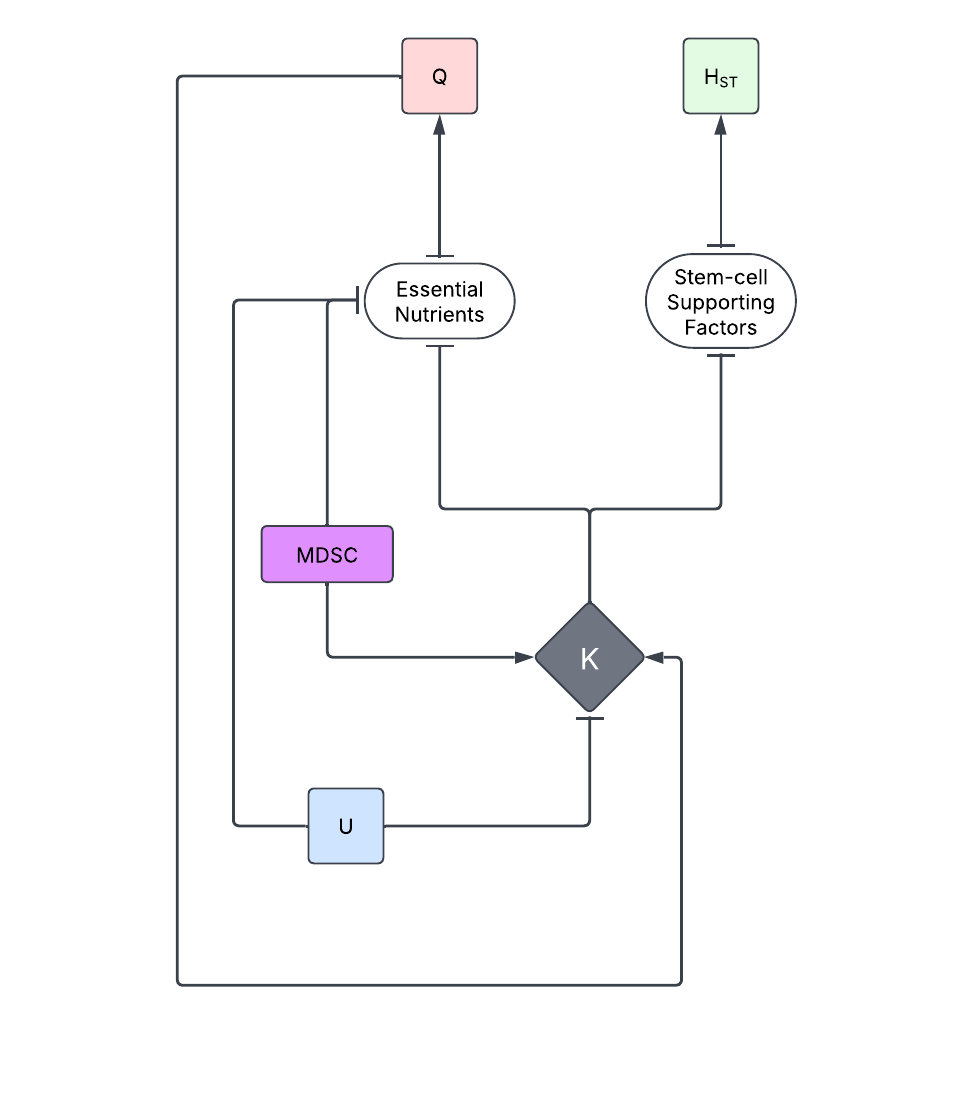}
  \caption{Tissue damage applies negative pressure on molecular factors and SCSFs which are necessary for maintaining activated pro-inflammatory WBCs and stem cell populations respectively; immuno-suppressive WBCs heal tissue damage}\label{fig:nodes_damage}
\end{figure}

Damage associated molecular patterns (DAMPs), are a class of endogenous molecules that are released by damaged or dying cells. They act as danger signals, signaling the presence of foreign microbes or some other source of damage, instigating the innate immune response \cite{roh2018damage,ma2024damps}. These DAMPs bind onto PRRs activating a range of different cellular signaling pathways, most notably \textit{TLR} and \textit{STAT3} signaling pathways, that can induce either a pro-inflammatory response or anti-inflammatory response depending on the specific cell encountered and cellular milieu. 

The dynamics for $K$ is defined to be
\begin{equation*}
    \frac{dK}{dt} = \beta_N N + S_{kd}\left[\frac{(\tau_QP + \tau_U A + k_{sn} N) \cdot \Psi S}{\tau_QP + \tau_U A + k_{sn} N + \Psi S} \cdot \frac{k_{sn} N}{\tau_Q P + \tau_U A + k_{sn} N}\right] + S_{kq} Q - \frac{R_{ku} U K}{R_{ku} U + K},
\end{equation*}
 where $\beta_N N$ represents the damage to surrounding non-specific, non-hematopoietic tissue by pathogens, $S_{kd}(\frac{(\tau_QP + \tau_U A + k_{sn} N) \cdot \Psi S}{\tau_QP + \tau_U A + k_{sn} N + \Psi S} \cdot \frac{k_{sn} N}{\tau_Q P + \tau_U A + k_{sn} N})$ represents immature hematopoietic cells interacting with and being damaged/killed by pathogen cells (thus releasing $S_{kd}$ DAMPs in the process), $S_{kq}Q$ represents the act of activated immune cells themselves damaging non-specific, non-hematopoietic tissue tissue, either through direct action or by the release of reactive oxygen species (ROS), which can cause oxidative stress and DNA damage, and $\frac{R_{ku} U K}{R_{ku} U + K}$ represents the ability of mature anti-inflammatory cells to heal tissue damage.

 The role of DAMPs in the immune response is complex, instigating both pro-inflammatory and anti-inflammatory responses. In our model, we choose to follow the following schema:
\begin{itemize}
    \item \textbf{Low tissue damage} $\Rightarrow$ Pro-inflammatory response is preferentially activated
    \item \textbf{High tissue damage} $\Rightarrow$ Anti-inflammatory response is preferentially activated
\end{itemize}
 This allows us to keep the model simple while producing qualitatively accurate behavior. Similar to how we modeled this for PAMPs, we construct the following functions, which are used in the crowding function \eqref{eq:crowding1}:
\begin{equation*}
\begin{split}
    \text{Amp}_{PK}(K) &= 0.5 \frac{K}{K + \theta_K} + 1,\\
    \text{Amp}_{AK}(K) &= \frac{K}{K + \theta_K} + 0.75.
\end{split}
\end{equation*}
 The specific values chosen for each (the coefficients of the michaelis-menten type function and the added value) where chosen semi-arbitrarily but to be consistent with the schema presented above. Specifically, we see that for values of tissue damage $K \leq \theta_K$, $\text{Amp}_{AK}(K) \leq \text{Amp}_{PK}(K)$ and conversely when $K \geq \theta_K$, $\text{Amp}_{AK}(K) \geq \text{Amp}_{PK}(K)$.

 We now define two related dynamical quantities for stem cell supporting factors ($SCSF$) and essential nutrients ($EN$). $SCSF$ is a compartment of molecules up-regulated by BM niche tissue which is essential for retention, survival, and proliferation of HSPCs. The $SCSF$ compartment includes factors such as the aptly named stem cell factor, Flt-3 ligand, and stromal derived factor-1 (also known as CXCL12) \cite{lapid2013egress}. We define $SCSF$ to be,
\begin{equation*}
    \frac{dSCSF}{dt} = S_{SCSF} \frac{K_{crit}}{K_{crit} + K} - d_{SCSF}SCSF.
\end{equation*}
 As before, the rightmost term is a standard exponential decay term. The leftmost term, $S_{SCSF} \frac{K_{crit}}{K_{crit} + K}$, induces a steady rate of SCSF secretion as a function of tissue damage. As tissue damage increases, the ability of the system to up-regulate the ingredients necessary for maintenance of a robust HSPC pool decreases. In other words, as $K \rightarrow \infty$, $S_{SCSF} \frac{K_{crit}}{K_{crit} + K} \rightarrow 0$.

 Similarly, $EN$ is a compartment of molecules composed of amino acids, vitamins, and minerals necessary for the proliferation and proper functioning of immune cells. Specifically, this compartment is linked with the survival and function of mature pro-inflammatory immune cells $Q$. The mechanics of this compartment are informed by molecules such as L-arginine, which is necessary for T-cell metabolism and anti-tumor properties \cite{geiger2016arginine}, and iron, which is known to regulate certain immunological processes \cite{cronin2019role,xiong2023nutrition}, to name a few.  We define $EN$ to satisfy,
\begin{equation*}
    \frac{dEN}{dt} = S_{EN} \frac{K_{crit}}{K_{crit} + K} - d_{EN} EN.
\end{equation*}
 Each term here plays the same role as its equivalent term in $\frac{dSCSF}{dt}$, and each variable plays a similar role to its respective compartments ($SCSF \leftrightarrow H_{ST}$ and $EN \leftrightarrow Q$) in that they each place an upper limit on them at any one time, with this upper limit being determined by how much damage the system has sustained $K$. 
 
 To account for the introduction of $K$, $EN$, and $SCSF$, we modify $H_{ST}$ and $Q$ to become
\begin{align*}
     \frac{dH_{ST}}{dt} &= \left[\Gamma +\frac{(\Delta-\Gamma) I}{H_{crit} + I}\right] H_{ST} \left[1 - \frac{C_{sh}H_{ST}}{SCSF}\right] - \frac{\alpha \cdot I \cdot P \cdot H_{ST}}{\alpha \cdot I \cdot H_{ST} + P},\\
     \frac{dQ}{dt} &= \frac{(\tau_QP + \tau_U A + k_{sn} N) \cdot \Psi S}{\tau_QP + \tau_U A + k_{sn} N + \Psi S} \cdot \frac{\tau_Q P}{\tau_Q P + \tau_U A + k_{sn} N} \cdot \frac{\frac{1}{2} EN}{\frac{1}{2} EN + C_{QE} Q + C_{UE} U} - d_Q Q,
\end{align*}
 where $C_{ij}$ is the rate at which the molecule \textit{i} is consumed per cell \textit{j}, and $\Gamma$ and $\Delta$ are the lower and upper bounds on $H_{ST}$ self-renewal rates, respectively. We model $H_{ST}$ self-renewal as a function of the pro-inflammatory force which is consistent with current understanding of stem cell dynamics. $SCSF$ has an additional effect on the mobilization of HSPCs ($H_{ST} \rightarrow MPP$) -- it counteracts the effects of pro-inflammatory molecules mobilizing HSPCs; adhesion molecules, such as the previously mentioned CXCL12, keep these  cells attached to the BM Niche tissue and in a \textit{G0} phase. To achieve this effect in our model, we construct another amplifying function
\begin{equation*}
    \text{Amp}_{ASC}(SCSF) = 0.75 \frac{SCSF}{\theta_K + SCSF} + 1,
\end{equation*}
 and define
\begin{equation*}
    I_{H}(P, A, N, K, SCSF) = \frac{\text{Amp}_{PN}(N) \cdot \text{Amp}_{PK}(K) \cdot P}{\text{Amp}_{PN}(N) \cdot \text{Amp}_{PK}(K) \cdot P + \text{Amp}_{AK}(K) \cdot \text{Amp}_{ASC}(SCSF) \cdot A},
\end{equation*}
 where we will often suppress the dependence of $I_H$ on $P, A, N, K, SCSF$ to reduce cluttered notation: $I_H \equiv I_H(P, A, N, K, SCSF)$. 
 
 Given that the positive effect on $H_{ST} \rightarrow MPP$ would only occur for the non-mobilized cells $H_{ST}$, this change necessitates the use of two different inflammation crowding terms used for $H_{ST}$ and $MPP$. The $MPP$ cells are already mobilized and have moved sufficiently out of the BM Niche for these molecules to not have the same effect. For $H_{ST}$ we use $I_{H}(P, A, N, K, SCSF)$ and for $MPP$ we will use $I(P, A, N, K)$. The two functions are nearly identical except the former utilizes $\text{Amp}_{ASC}(SCSF)$ whereas the latter does not, thus the updated derivatives for each are

\begin{align*}
    \frac{dH_{ST}}{dt} &= \Gamma (\frac{I_{H}}{H_{crit} + I_{H}}) (\Delta-\Gamma) H_{ST} \left[1 - \frac{C_{sh}H_{ST}}{SCSF}\right] - \frac{\alpha \cdot I_H \cdot P \cdot H_{ST}}{\alpha \cdot I_H \cdot H_{ST} + P},\\
    \frac{dMPP}{dt} &= \frac{\alpha \cdot I_H \cdot P \cdot H_{ST}}{\alpha \cdot I_H \cdot P + H_{ST}} - \frac{\alpha \cdot P \cdot I \cdot MPP}{\alpha \cdot P \cdot I + MPP} - d_H MPP
\end{align*}

\subsubsection{MDSCs}\label{si:mdsc}

\begin{figure}[ht!]
    \centering
      \includegraphics[width=\textwidth]{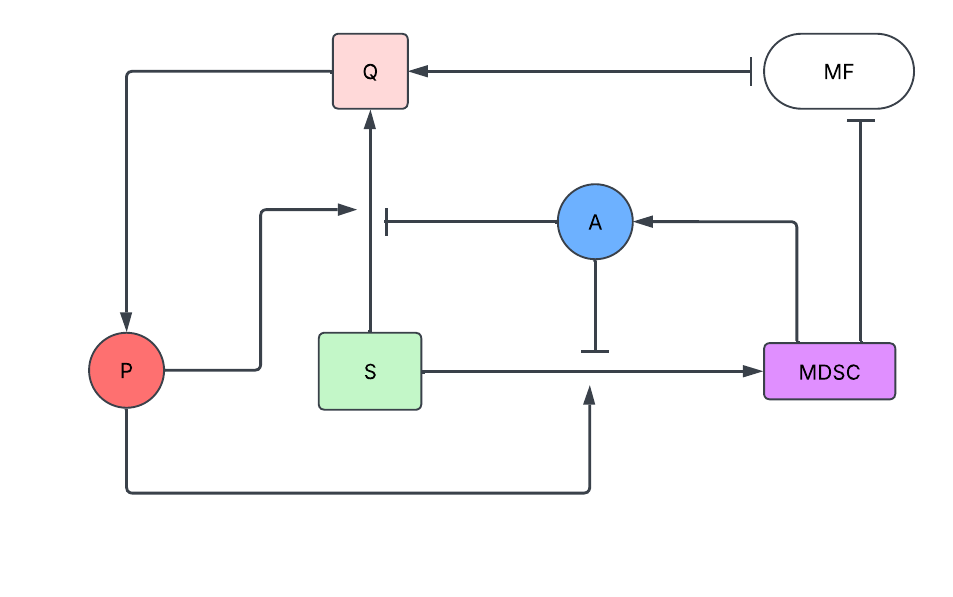}
      \caption{\textit{MDSCs} are up-regulated by pro-inflammatory signals, suppress \textit{Q} by targeting their essential nutrients and secreting anti-inflammatory signals }\label{fig:nodes_mdsc}
\end{figure}

\textbf{MDSCs}, which stands for \textit{myeloid-derived suppressor cell}, is a name used to refer to a heterogeneous group of immature myeloid cells with potent immuno-suppressive abilities \cite{veglia2021myeloid} \cite{gabrilovich2009myeloid}. These cells do not appear in large quantities during steady state and serve an important function in suppressing the immune response during pathologic conditions to prevent excessive pro-inflammatory responses that can damage healthy tissue \cite{gabrilovich2017myeloid}. These cells can appear in large quantities during states of chronic low-grade infection where their development into terminal, mature myeloid WBCs (such as neutrophils and macrophages) is arrested and they remain in an immature, immuno-suppressive phenotype. The exact set of mechanisms that promotes the expansion of these cells is still under study, though significant progress has been made in recent years. Specifically, for this model we are interested in the TLR and STAT3 signaling pathways in MDSCs that are activated by PAMPs and certain pro-inflammatory cytokines, respectively \cite{gabrilovich2009myeloid}, and which, when activated, induce changes in the MDSC cells that promote the production of inducible nitric oxide synthase (iNOS), which reacts with L-arginine to produce nitric oxide which is known to play a role in immune modulation, pathogen control, and tissue damage in high concentrations \cite{sharma2007role} \cite{botta2008anti} \cite{abramson2001role}; Arginase, which is known for catalyzing the hydrolysis of L-arginine which starves T-cells of this amino acid essential for their functioning \cite{rodriguez2007arginine} \cite{caldwell2018arginase}; and TGF-$\beta$, which is a pleiotropic cytokine that is well known for its anti-inflammatory effects \cite{sanjabi2017regulation}. 
\newline
\newline
From a modeling perspective, the role of \textit{MDSCs} and the mature, immuno-suppressive WBC compartment \textit{U} appear to be extremely similar in that they act as a check on the pro-inflammatory response to prevent excessive activation which can lead to tissue damage and eventually death. The key difference is that MDSCs are up-regulated by the same signals that up-regulate the pro-inflammatory response itself, \textit{P}. This simple distinction is crucial as this means that \textit{MDSCs} act as a first-response check on pro-inflammation while \textit{U} cells expand during the resolution phase of acute inflammation. To model this, we start with the following equation

\begin{align*}
    \frac{dMDSC}{dt} &= (1 - \Omega)\left[\frac{(\tau_Q P + \tau_U A + k_{sn} N) \Psi S}{\tau_Q P + \tau_U A + k_{sn} N + \Psi S}\right]\left[\frac{\tau_Q P}{\tau_Q P + \tau_U A + k_{sn} N}\right] \\
    & \quad  - d_M \left[\frac{0.5  Q^k}{0.5 MDSC^k + Q^k} + 0.5\right] MDSC,
\end{align*}
 where $0 \leq \Omega \leq 1$ is a parameter controlling the proportion of total interactions $S + P \rightarrow Q, \; MDSC$ goes to $MDSC$ (e.g. if $\Omega = 0.4$ and the total rate of $(S+P) = (\frac{(\tau_Q P + \tau_U A + k_{sn} N) \Psi S}{\tau_Q P + \tau_U A + k_{sn} N + \Psi S})(\frac{\tau_Q P}{\tau_Q P + \tau_U A + k_{sn} N}) = 100$, then the variable \textit{MDSC} would increase by $0.4 \; \times \; 100 = 40$ i.e. we expect to see a gain of $40$ cells in the compartment and conversely $Q$ would gain $0.6 \; \times \; 100 = 60$ cells).

 To the right we utilize another exponential decay term albeit with a modification in the form of $( \frac{0.5 Q^k}{0.5 MDSC^k + Q^k} + 0.5)$. Observe that as $Q \rightarrow \infty$, this term $\rightarrow 1$ and as $MDSC \rightarrow \infty$, this term $\rightarrow 0.5$. The formulation of this term is informed by recent findings supporting the idea that activated T-cells can retaliate against MDSCs by suppressing the immuno-suppressive functions of MDSCs as well as inducing apoptosis in this population \cite{chen2021reprogramming}, and it is this latter point which the above term represents. As before, the specific values of the coefficient of the hill-type function and its additive, both of which are set to $0.5$ here, are largely arbitrary. The only constraint is that these two values are between 0 and 1 and that they add up to 1. Specific values were set so as to prevent having to add anymore parameters than necessary.

 Another distinguishing factor between \textit{U} and \textit{MDSCs} lies in \textit{U}'s ability to both heal tissue damage (implemented as $ - \frac{R_{ku} U K}{R_{ku} U + K}$ in $\frac{dK}{dt}$) and \textit{U}'s ability to directly down-regulate the pro-inflammatory signal population (implemented as $-C_{UP} U$ in $\frac{dP}{dt}$), whereas \textit{MDSCs} can actually increase tissue damage by secretion of tissue-damaging ROS molecules which contributes to their immuno-suppressive ability \cite{huang2023function}. To reflect this, we modify the equation for tissue damage as follows 
\begin{equation*}
    \begin{split}
    \frac{dK}{dt} = \beta_N N + S_{kd}(\frac{(\tau_QP + \tau_U A + k_{sn} N) \cdot \Psi S}{\tau_QP + \tau_U A + k_{sn} N + \Psi S} \cdot \frac{k_{sn} N}{\tau_Q P + \tau_U A + k_{sn} N}) + S_{kq} Q + \\ S_{km}MDSC (\frac{P + A + N}{1/2 \cdot MDSC + P + A + N}) - \frac{R_{ku} U K}{R_{ku} U + K}.
    \end{split}
\end{equation*}

 With \textit{MDSCs} now established as a state variable, we modify the equations for $Q$ and $A$ as follows
\begin{equation*}
    \frac{dQ}{dt} = \Omega \cdot \frac{(\tau_QP + \tau_U A + k_{sn} N) \cdot \Psi S}{\tau_QP + \tau_U A + k_{sn} N + \Psi S} \cdot \frac{\tau_Q P \cdot Amp_{PN} \cdot Amp_{PK}}{\tau_Q P + \tau_U A + k_{sn} N} \cdot \frac{\frac{1}{2} EN}{\frac{1}{2} EN + C_{QE} Q + C_{UE} U + C_{ME} MDSC} - d_Q Q 
\end{equation*}

\begin{equation*}
    \frac{dA}{dt} = S_{AM}MDSC + S_{AH}MPP + S_{AS}S + S_{AU}U - d_A A 
\end{equation*}

\begin{figure}[ht!]
\centering
  \includegraphics[width=\textwidth]{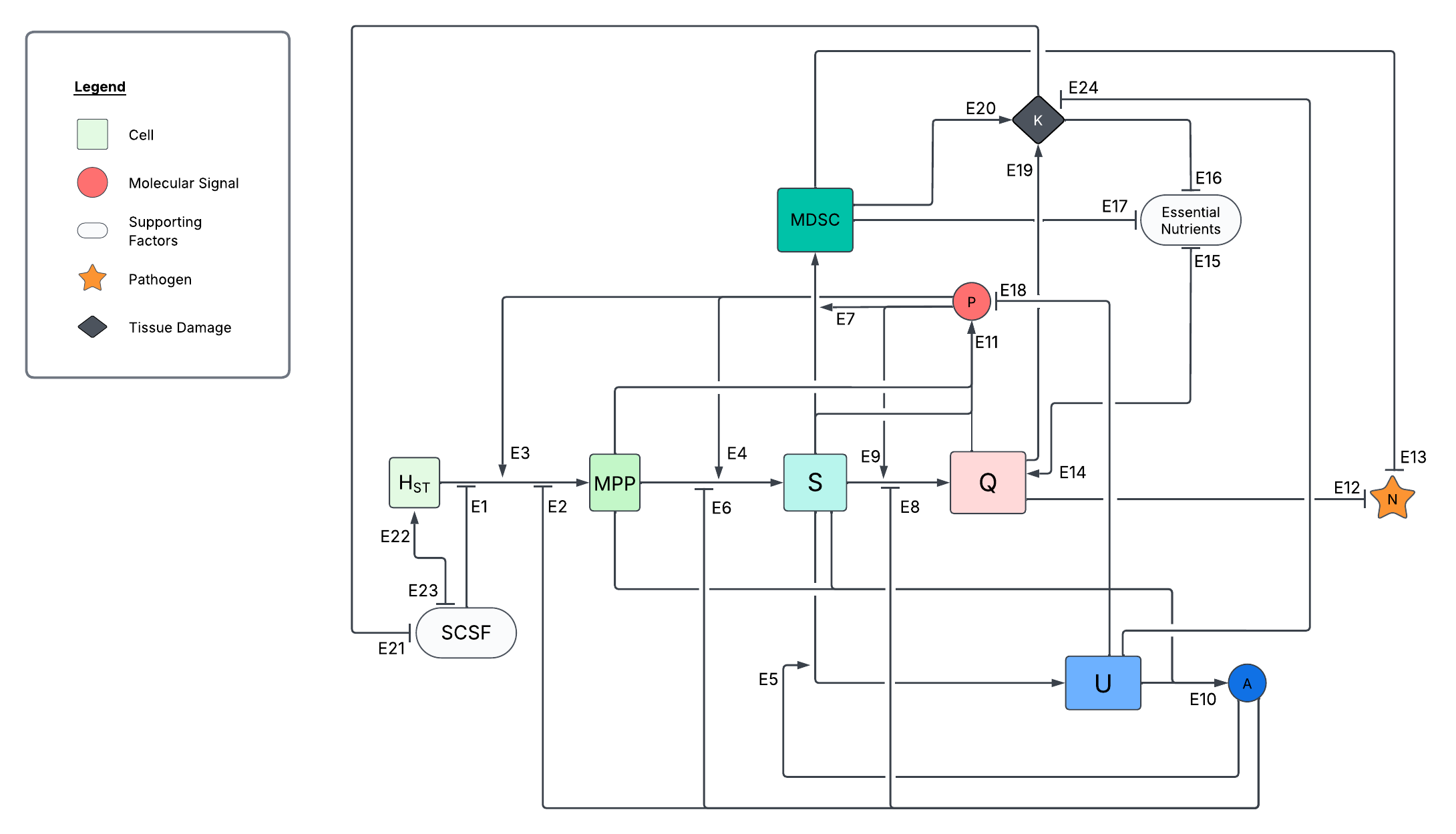}
  \caption{An overview of our 12-variable hematopoiesis and inflammation model. White ovals represent molecular nutrients necessary for supporting cellular populations, rectangular nodes represent compartments of cells, and circles represent molecular signals (in the form of pro- and anti-inflammatory cytokines). Pathogens are denoted by a star, and tissue damage is denoted by a black diamond. The labels E1 through E24 are detailed in Appendix \ref{si:model}.}\label{fig:nodes}
\end{figure}

\subsection{Model Equations}\label{sec:model_equations}

 Our model contains the following variables (see Appendix Section \ref{si:model} for a detailed description of each variable):
\begin{itemize}
\item Cells:
\begin{itemize}
    \item $MDSC$: Myeloid-derived suppressor cell.
    \item $H_{ST}$: Short-term HSPCs.
    \item $MPP$: Multipotent progenitor (MPP) cells.
    \item $S$: Immature WBCs.
    \item $Q$, $U$: Pro- and anti-inflammatory WBCs, respectively.
\end{itemize}
\item Molecular signals:
\begin{itemize}
    \item $P$, $A$: Pro- and anti-inflammatory cytokines, respectively.
\end{itemize}
\item Supporting factors:
\begin{itemize}
    \item $EN$: Essential Nutrients.
    \item $SCSF$: Stem cell supporting factors.
\end{itemize}
\item Other Variables:
\begin{itemize}
    \item $K$: Damage Associated Molecular Patterns (DAMPs).
    \item $N$: Pathogens.
\end{itemize}
\end{itemize}

 The model equations are as follows (see Appendix Section \ref{si:model} for a detailed derivation of all equations, and Appendix Section \ref{si:parameters} for all model parameters).
 
 Cellular dynamics are given by,
\begin{equation}\label{eq:model_a}
    \begin{split}
    \frac{dMDSC}{dt} &= (1 - \Omega)\frac{(\tau_Q P + \tau_U A + k_{sn} N) \Psi S}{\tau_Q P + \tau_U A + k_{sn} N + \Psi S}\cdot \frac{\tau_Q P \cdot \text{Amp}_{PN}(N) \cdot \text{Amp}_{PK}(K)}{I_S} \\
        &\quad\quad - d_M \left[ \frac{0.5 Q^k}{0.5 MDSC^k + Q^k} + 0.5\right] MDSC,\\
    \frac{dH_{ST}}{dt} &= \left[\Gamma + \frac{I(\Delta - \Gamma)}{I + H_{crit}}\right] H_{ST} \left[1 - \frac{C_{hs}H_{ST}}{SCSF}\right] - \frac{\alpha \cdot I_{H} \cdot P \cdot H_{ST}}{\alpha \cdot I_{H} \cdot P + H_{ST}},\\
    \frac{dMPP}{dt} &= \frac{\alpha \cdot I_{H} \cdot P \cdot H_{ST}}{\alpha \cdot I_{H} \cdot P + H_{ST}} - \frac{\alpha \cdot P \cdot I \cdot MPP}{\alpha \cdot MPP \cdot I + P} - d_H MPP,\\
    \frac{dS}{dt} &= \frac{\alpha \cdot P \cdot I \cdot MPP}{\alpha \cdot MPP \cdot I + P} \cdot \left[1 + \frac{I}{I + I_{\text{crit}}}\right] - \frac{(\tau_QP + \tau_U A + k_{sn} N) \cdot \Psi S}{\tau_QP + \tau_U A + k_{sn} N + \Psi S} - d_S S,\\
    \frac{dQ}{dt} &= \Biggl[\Omega \cdot \frac{(\tau_QP + \tau_U A + k_{sn} N) \cdot \Psi S}{\tau_QP + \tau_U A + k_{sn} N + \Psi S} \cdot \frac{\tau_Q P \cdot \text{Amp}_{PN}(N) \cdot \text{Amp}_{PK}(K)}{I_S} \\
    &\quad\quad\quad\quad\quad\quad \cdot \frac{0.5 EN}{0.5 EN + C_{QE} Q + C_{UE} U + C_{ME} MDSC}\Biggr] - d_Q Q,\\
    \frac{dU}{dt} &= \frac{(\tau_QP + \tau_U A + k_{sn} N) \cdot \Psi S}{\tau_QP + \tau_U A + k_{sn} N + \Psi S} \cdot \frac{\tau_U A \cdot \text{Amp}_{AK}(K)}{I_S} - d_U U,
\end{split}
\end{equation}
where $I \equiv I(P,A,N,K)$, $I_H \equiv I_H(P,A,N,K,SCSF)$, and $I_S \equiv I_S(P,A,N)$ (the functions $I$, $I_H$, and $I_S$ are defined below in Equation \eqref{eq:I_functions}). Molecular signals are given by,
\begin{equation}\label{eq:model_b}
    \begin{split}
        \frac{dP}{dt} &= (S_{PH}MPP + S_{PQ}Q)(0.8 \; I + 0.2) + S_{PS}S - d_P P - \frac{C_{UP} U (1 - d_P) P}{C_{UP} U + (1- d_P) P},\\
    \frac{dA}{dt} &= S_{AM}MDSC + S_{AH}MPP + S_{AS}S + S_{AU}U - d_A A,
    \end{split}
\end{equation}
supporting factors are given by,
\begin{equation}\label{eq:model_c}
    \begin{split}
        \frac{dEN}{dt} &= S_{EN} \frac{K_{crit}}{K_{crit} + K} - d_{EN} EN,\\
    \frac{dSCSF}{dt} &= S_{SCSF} \frac{K_{crit}}{K_{crit} + K} - d_{SCSF}SCSF,
    \end{split}
\end{equation}
and the remaining variables (pathogens and DAMPs) satisfy,
\begin{equation}\label{eq:model_d}
    \begin{split}
        \frac{dN}{dt} &= g_N N \left[1 - \frac{N}{N_{\infty}}\right] - \frac{(k_{nq}Q + k_{ns}S) N}{N + N_{1/2}}+ S_{kq} Q + S_{km}MDSC - \frac{R_{ku} U K}{R_{ku} U + K},\\
        \frac{dK}{dt} &= \beta_N N + S_{kd}\left[\frac{(\tau_QP + \tau_U A + k_{sn} N) \cdot \Psi S}{\tau_QP + \tau_U A + k_{sn} N + \Psi S} \cdot \frac{k_{sn} N}{\tau_Q P + \tau_U A + k_{sn} N}\right].
    \end{split}
\end{equation}
 Parameters are as follows: $S_{ij}$ is the rate at which molecule $i$ is secreted \textit{per} unit of cell $j$, $C_{ij}$ is the rate at which the molecule $i$ is consumed per cell $j$, $k_{nj}$ represents the kill rate of pathogens $N$ per cell $j$, $g_j$ represents the effective growth rates for cell or species $j$, and $d_j$ represents the effective death rates for cell or species $j$.

 The remaining functions are given by,
\begin{equation}\label{eq:I_functions}
\begin{split}
    I(P, A, N, K) &= \frac{\text{Amp}_{PN}(N) \cdot \text{Amp}_{PK}(K) \cdot P}{\text{Amp}_{PN}(N) \cdot \text{Amp}_{PK}(K) \cdot P + \text{Amp}_{AK}(K) \cdot A},\\
    I_{H}(P, A, N, K, SCSF) &= \frac{\text{Amp}_{PN}(N) \cdot \text{Amp}_{PK}(K) \cdot P}{\text{Amp}_{PN}(N) \cdot \text{Amp}_{PK}(K) \cdot P + \text{Amp}_{AK}(K) \cdot \text{Amp}_{ASC}(SCSF) \cdot A}\\
    I_S(P, A, N) &= \tau_Q P \cdot \text{Amp}_{PN}(N) \cdot \text{Amp}_{PK}(K) \; + \; \tau_U A \cdot \text{Amp}_{AK}(K) \; + \; k_{sn} N,
\end{split}
\end{equation}
where
\begin{equation*}
\begin{split}
    \text{Amp}_{PN}(N) &= \frac{N^k}{N^k + \theta_N} + 0.25\\
    \text{Amp}_{PK}(K) &= 0.5 \frac{K}{K + \theta_K} + 1,\\
    \text{Amp}_{AK}(K) &= \frac{K}{K + \theta_K} + 0.75,\\
    \text{Amp}_{ASC}(SCSF) &= 0.75 \frac{SCSF}{\theta_K + SCSF} + 1.
\end{split}
\end{equation*}
 $k$ is a hill-type coefficient and $\theta_N$ or $\theta_K$ represents the concentration at which this reaches half of its maximum amplitude. The constants used for each term were chosen to reflect the effect at the steady-state level in the absence of pathogens or damage. For example, a value of $0.25$ implies that, in the absence of pathogens, stimulating the pathogen receptors (including toll-like receptors (TLR) and pattern recognition receptors (PRR)) dampens the pro-inflammatory force to a quarter of its strength. Specifically, we see that for values of tissue damage $K \leq \theta_K$, $\text{Amp}_{AK}(K) \leq \text{Amp}_{PK}(K)$ and conversely when $K \geq \theta_K$, $\text{Amp}_{AK}(K) \geq \text{Amp}_{PK}(K)$. More detailed descriptions are provided in Appendix Section \ref{si:damage}. Each $I$ function represents an \textit{inflammation metric}, a proxy for measuring the net inflammatory force acting on specific subgroups of cell populations, while $Amp_{ij}(...)$ stand for amplifier functions and imply that force $i$ (either $P$ or $A$ for pro-inflammatory / anti-inflammatory force respectively) is amplified by force $j$.

 This model qualitatively recreates leukopoiesis under steady-state health conditions and under infection-driven inflammatory processes (see Appendix Section \ref{si:model} for a detailed derivation of all equations, and Appendix Section \ref{si:parameters} for all model parameters).

\section{Results}\label{sec:results}

    \begin{figure}[ht!]
        \includegraphics[width=\textwidth]{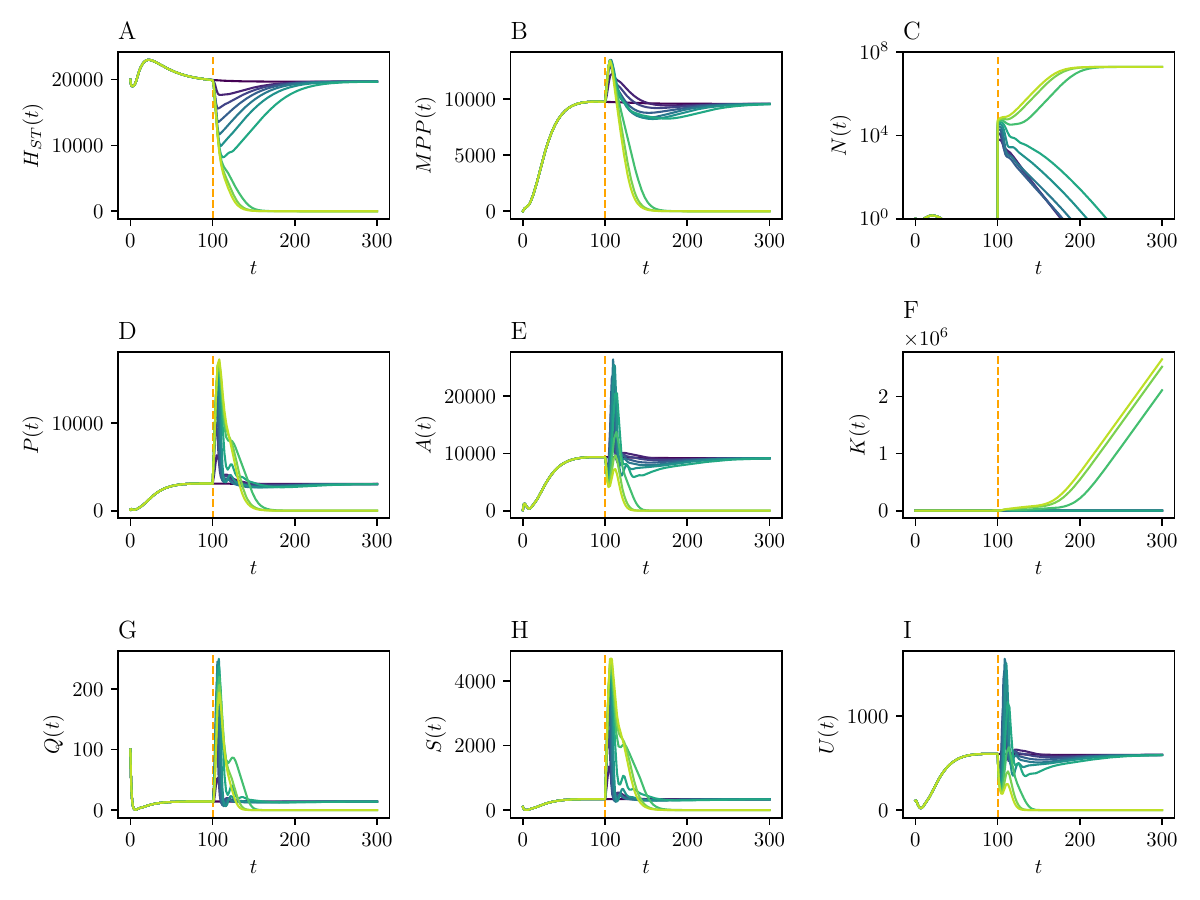}
        \caption{Recovery from sepsis. Solution curves correspond to different intensities of pathogen insults. The insults range from $N(100) = 0$ (purple) to $N(100) = 54,000$ (light green). The vertical dashed orange line at $t=100$ is the time at which the pathogenic insult is applied. Note that the y-axis of Panel C is on a logarithmic scale. A: Short-term HSPCs ($H_{ST}$), B: Multipotent progenitor cells ($MPP$), C: Pathogens ($N$), D: Pro-inflammatory cytokines ($P$), E: Anti-inflammatory cytokinds ($A$), F: Damage associated molecular patterns ($K$), G: Pro-inflammatory WBCs ($Q$), H: Immature WBCs ($S$), I: Anti-inflammatory WBCs ($U$).  See Tables \ref{tab:parameters}, \ref{tab:nominal_params}, and \ref{tab:additional_parameters}  in Appendix \ref{si:parameters} for model parameter values and Table \ref{tab:hyper} for simulation parameters.}\label{fig:recovery}
    \end{figure}

    \subsection{Model Replicates Clinical Trajectories}\label{sec:model_replicates}
    We aimed to observe the following qualitative behaviors that resemble those seen in patients who undergo systemic infection-related health complications:
    \begin{enumerate}
        \item \textbf{Recovery} and \textbf{Septic Death}  (Figure \ref{fig:recovery}).
        \item Secondary (\textbf{Nosocomial}) Infection (Figure \ref{fig:nosocomial}).
        \item \textbf{Chronic Infection}  (Figure \ref{fig:chronic}).
        \item \textbf{Aseptic Death} (Figure \ref{fig:aseptic}).
    \end{enumerate}

    \subsubsection{Recovery and Septic Death}

    In Figure \ref{fig:recovery}, we show model solutions after applying a range of 9 pathogenic insults at $t=100$ (ranging from $N(100)=0$ (purple) to $N(100)=54,000$ (light green). In each panel, each of the 9 trajectories corresponds to a single simulation and a single pathogenic insult value. For pathogenic insults below a threshold of approximately $N=18,000$, $H_{ST}$ returns to pre-infection values (recovery). For pathogenic insults above this threshold, we observe $N(t) \rightarrow N_{\infty}$ and $H_{ST}(t) \rightarrow 0$. This behavior corresponds to a runaway infection that ultimately leads to death in a patient (septic death). In all cases, the model successfully demonstrates the tendency of the pro-inflammatory compartments to increase during the onset phase of inflammation and their subsequent decline and the rise of the anti-inflammatory compartments during the resolution phase of infection.

    \subsubsection{Nosocomial Infection}
    
\begin{figure}[ht!]
        \includegraphics[width=\textwidth]{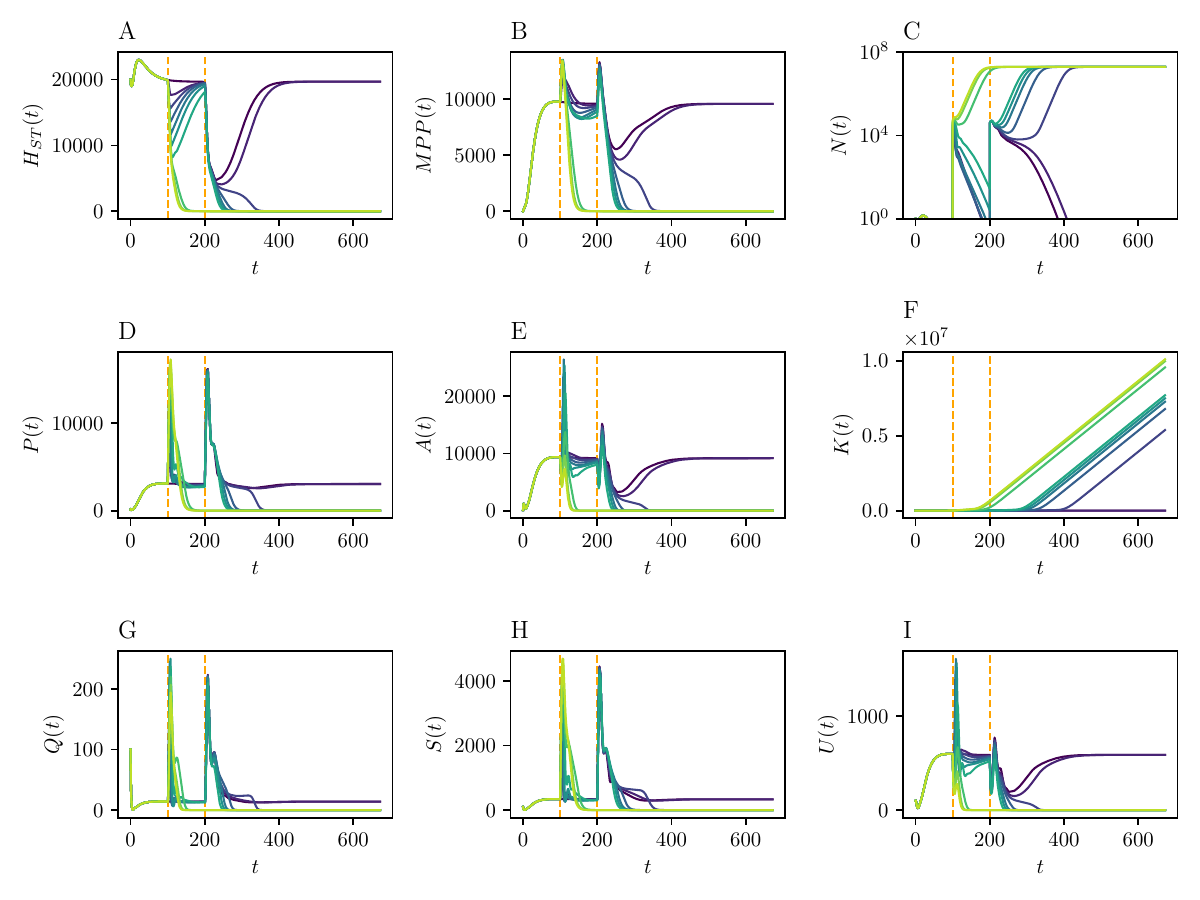}
        \caption{Nosocomial (secondary) infection. These simulations are identical to Figure \ref{fig:recovery} except for an additional, secondary infection at $t=200$. See Tables \ref{tab:parameters}, \ref{tab:nominal_params}, and \ref{tab:additional_parameters}  in Appendix \ref{si:parameters} for model parameter values and see Table \ref{tab:hyper} for simulation hyper parameters.}\label{fig:nosocomial}
    \end{figure}

    In Figure \ref{fig:nosocomial}, we show model solutions after applying the same range of 9 pathogenic insults, but now include a secondary (nosocomial) insult at $t=200$. Some solutions that were able to return to a healthy steady state in Figure \ref{fig:recovery} fail to do so after the nosocomial infection is introduced. The parameter values were kept the same as in Figure \ref{fig:recovery}. These simulations show that a secondary infection can lead to septic death, even when a patient may have recovered otherwise.

    % $K > 0$, an initial infection reduces the health of the system enough that the introduction of a secondary infection leads to $N \rightarrow N_{\infty}$

    \begin{table}[ht!]
    \centering
    \caption{Simulation hyper parameters for Figures \ref{fig:recovery} and \ref{fig:nosocomial}}\label{tab:hyper}
    \begin{tabular}{|c|c|c|}
        \hline
         \textbf{Condition} & \textbf{Definition} & \textbf{Value} \\ \hline
        Runs & How many solutions to compute & 10 \\ \hline
        Pathogen Default & Pathogen insult applied to first run & 0 \\ \hline
        Pathogen Increment & How much to increment the pathogen insult per run & 6000 \\ \hline
        Initial Time & Time of initial pathogen insult (pathogen occurs at $t=\text{Initial Time})$ & 100 \\ \hline
        Nosocomial Size & Magnitude of the secondary infection (this value is used for every simulation) & 40,000 \\ \hline
        Nosocomial Time & Time of secondary infection (occurs at $t=\text{Nosocomial Time}$) & 200 \\ \hline
    \end{tabular}
    \end{table}

    \subsubsection{Chronic Infection}

\begin{figure}[ht!]
        \includegraphics[width=\textwidth]{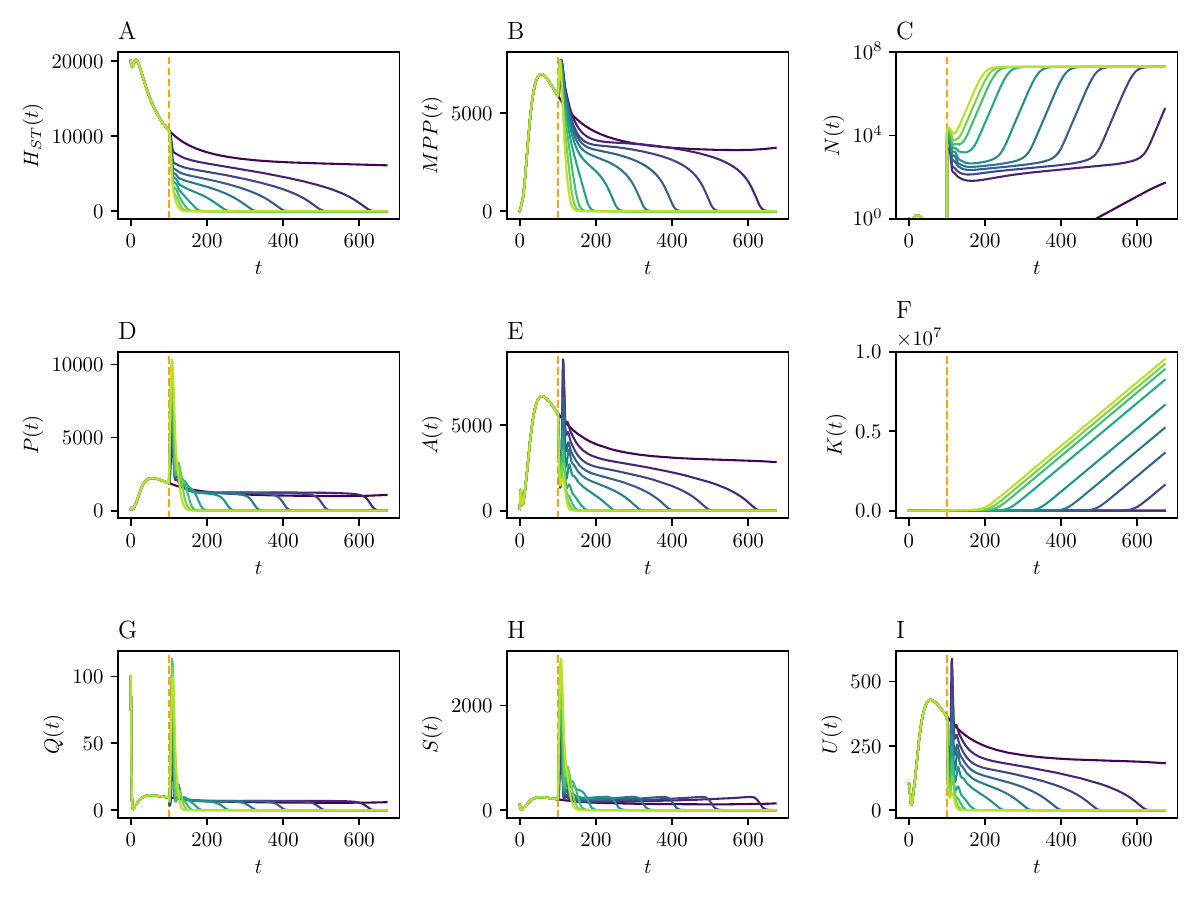}
        \caption{Chronic infection. A small initial infection of $N(0)=1$ is introduced, which alters long-term disease dynamics. The initial infection is subdued but not removed, which causes continuous tissue damage following a pathogenic insult at $t=100$ ranging from $N(0)=0$ (purple) to $N(0)=9000$ (light green). The system experiences a delayed septic death. See Tables \ref{tab:parameters}, \ref{tab:nominal_params}, and \ref{tab:additional_parameters} in Appendix \ref{si:parameters} for model parameter values. See Table \ref{fig:chronic} for simulation parameters. The parameter values are kept the same as in Figure \ref{fig:recovery} except for  $H_{crit}$: 0.2 (Recovery) $\rightarrow$ 0.5 (Chronic) and $k_{sn}$: 3 (Recovery) $\rightarrow$ 10 (Chronic).}\label{fig:chronic}
    \end{figure}

    In Figure \ref{fig:chronic}, we show simulations in which a low-grade infection persists in a patient (chronic infection). Chronic infection leads to a pathologically activated immune response and carries the risk of increased susceptibility to nosocomial infections. Mathematically, we define chronic infection as some nontrivial range of time $t$ such that $N_{min} \leq N(t) \leq N_{max}$, $|\frac{dN}{dt}| < \epsilon$ for chosen $N_{min}, \: N_{min}, \: \epsilon > 0$. The parameter values are kept the same as in Figure \ref{fig:recovery} except for  $H_{crit}$: 0.2 (Recovery) $\rightarrow$ 0.5 (Chronic) and $k_{sn}$: 3 (Recovery) $\rightarrow$ 10 (Chronic).

    Increasing $H_{crit}$ results in desensitized short-term HSPCs requiring larger pro-inflammatory stimuli to induce increased cell-cycling (important for sustaining cellular populations during acute inflammation / infection). Increasing $k_{sn}$ results in immature white blood cells that are more effective at removing pathogens. The model depicts a scenario in which the system falsely appears stable for long periods of time as the immature WBC population is strong enough to reduce the initial pathogen population to prevent uncontrolled growth but not enough to remove them entirely. Because the pathogen population is reduced, the pro-inflammatory force (Figure \ref{fig:chronic} Panels D and C) is not strong enough to induce increased self-renewal rate in the desensitized short-term HSPCs, leading to an underwhelming anti-inflammatory phase of the acute immune response (Figure \ref{fig:chronic} Panels E and I) which fails to resolve tissue damage (Figure \ref{fig:chronic} Panel F). This induces a positive feedback loop: increased tissue damage leads to a weaker hematopoietic system, which leads to a greater pathogen force, which leads to more tissue damage, ultimately resulting in delayed septic death. %These results suggest one possible mechanism by which chronic infection / delayed septic death may occur.
    
    \begin{table}[ht!]
    \centering
    \caption{Simulation hyper parameters for Figure \ref{fig:chronic}}
    \begin{tabular}{|c|c|c|}
        \hline
         \textbf{Condition} & \textbf{Definition} & \textbf{Value} \\ \hline
        Runs & How many solutions to compute & 10 \\ \hline
        Pathogen Default & Pathogen insult applied to first run & 0 \\ \hline
        Pathogen Increment & How much to increase the pathogen insult per run & 2000 \\ \hline
        Initial Time & Time of initial pathogen insult (pathogen occurs at $t=\text{Initial Time})$ & 100 \\ \hline
    \end{tabular}
    \end{table}
 %This is not an exhaustive list of all the conditions / parameter permutations that can produce behavior resembling chronic infection. Instead, we wish to highlight the system's ability to produce a chronic infection phenotype.

     \subsubsection{Aseptic Death}

\begin{figure}[ht!]
    \includegraphics[width=\textwidth]{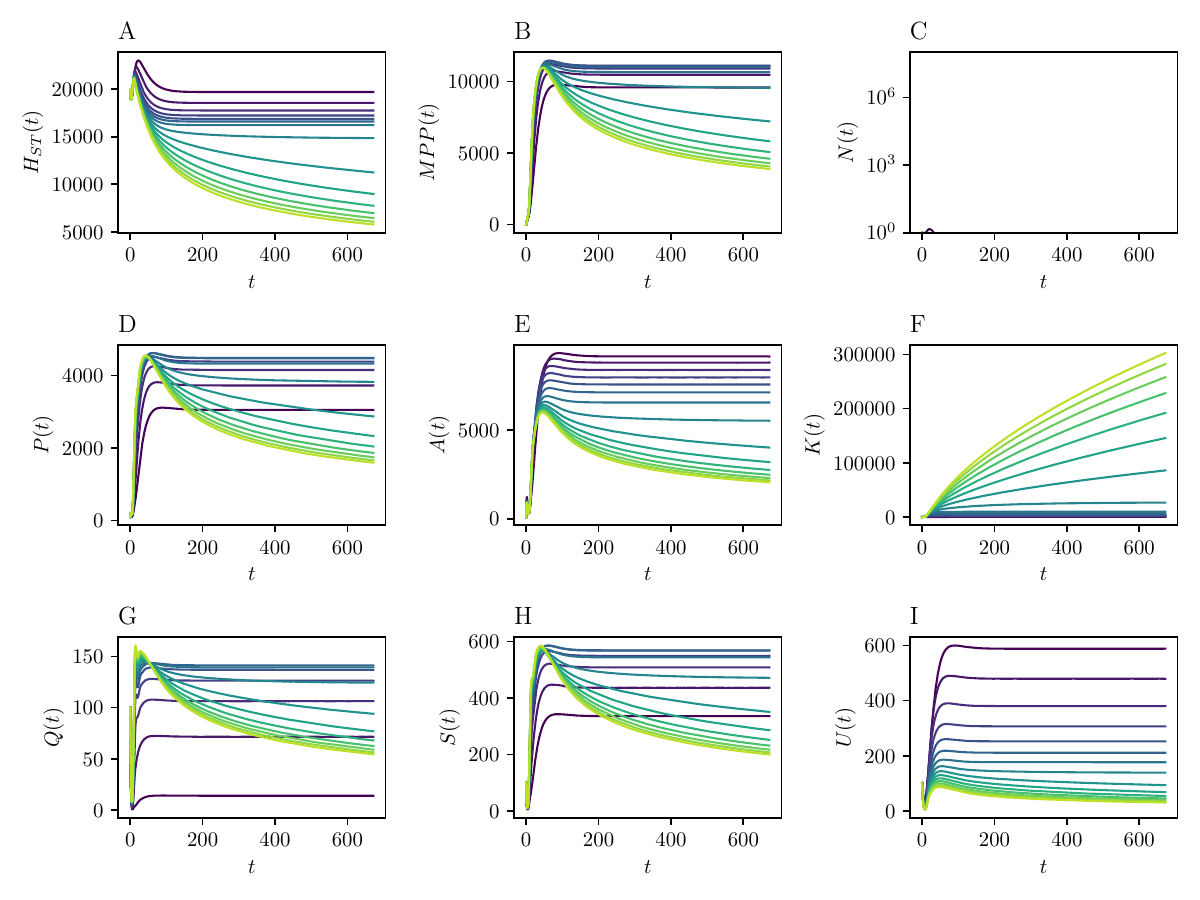}
    \caption{Aseptic death; purple corresponds to a normal steady state system and each run sees the system progressively becoming more weighted towards a pro-inflammatory phenotype leading to continuous tissue damage and reduced system health. The first solution (dark purple) corresponds to a ratio of $1:1$, the second solution to a ratio of $5:1$, and each solution after (3 and up) successively increment the ratio by $5:1$ (e.g. solution 3 uses a ratio of $10:1$). All other parameters are the same as Figure \ref{fig:recovery}.}\label{fig:aseptic}
\end{figure}  
 In Figure \ref{fig:aseptic}, we show simulations of aseptic death where we introduce no pathogenic insult. Instead, we alter the ratio between the two parameters $\tau_Q$ and $\tau_U$ with $\tau_U$ being held at a constant value of 1. Increasing this ratio corresponds to weighting immature white blood cell differentiation in favor of pro-inflammatory line differentiation (favoring mature pro-inflammatory WBCs and MDSC upregulation over anti-inflammatory mature WBCs). The increased ratios result in successively lower anti-inflammatory WBCs (Panel I) and successively higher pro-inflammatory WBCs (Panel G). This skews the balance between the pro-inflammatory and anti-inflammatory forces (Panels D and E, respectively) towards pro-inflammation. Past some currently undetermined critical threshold, this results in continuous tissue damage in the absence of pathogens (Panel F). The figure displays solutions which appear to stabilize and some which continually sustain heavy tissue damage up to the end of the digital runtime, but it is unknown if this particular metric ($\tau_Q : \tau_U$) is guaranteed to induce a non-zero stable fixed point in tissue damage for all values.

 Much like was the case with chronic infection, this is not an exhaustive exploration of all conditions which can tip the system into a chronic aseptic inflammation phenotype.

%$N = 0$, $\frac{dK}{dt} > 0$; this behavior corresponds to a state in which a patient's pro-inflammatory response is pathologically activated and tissue damage continues to occur without an external stimuli inducing it and carries the risk of increased susceptibility to nosocomial infections 

% \subsection{Parameter Scan}
% \ypa{I'll take care of this section - YP}

\subsection{Sensitivity Analysis}\label{sec:sensitivity}

 We performed sensitivity analysis on the model to identify key drivers of different model behaviors. We begin with the Morris sensitivity test \cite{Morris1991factorial} instead of Sobol because of the large number of model inputs we wanted to evaluate. While the Morris sensitivity test lacks the ability to precisely quantify the effects of input interactions, it requires far fewer model evaluations than Sobol. It is for this reason that Morris is often used as an initial screening followed by a more precise Sobol test on a reduced parameter space \cite{garcia2019robust}. 

 We provide a brief description of the Morris sensitivity test and its output metrics. The Morris sensitivity test, also known as the Elementary Effects test, is a type of global sensitivity test that examines the effects of each parameter on the model outputs. For a given model input parameter, the Morris method returns 3 metrics:
 \begin{enumerate}
     \item $\mu_i$: The average elementary effect of a model input parameter: it measures how much, on average, a model output changes in response to a change in the specified input parameter.
     \item $\mu_i^*$: The average absolute elementary effect: a useful way to study the absolute effect of a parameter input on model outputs, because it takes into account positive and negative changes, instead of simply considering an average. This is useful in cases where there are cancellation effects due to equal amounts of increase and decrease in the model output. 
     \item $\sigma_i$: The standard deviation of $\mu$. $\sigma_i$ provides a way of measuring the degree to which an input parameters interacts with other model parameters, and the degree to which the input parameter has nonlinear effects on model outputs.
 \end{enumerate}
 In our results below, we display each model parameter's $\mu_i^*$ and $\sigma$ scores to study which parameters influence chronic infection duration the most and whether these effects are linear or not.

 \subsubsection{Identification of key drivers of chronic infection}

 To assess chronic infection behavior over time, we tracked and assigned the total duration of time during which the system meets the requirements for \textit{chronic infection} (as defined in Section \ref{sec:model_replicates}). We used this time duration as the quantity of interest when performing the Morris sensitivity analysis.
 
 We performed 3 Morris sensitivity tests on nearly the entire parameter space to gain broad insight into which parameters may hold the most influence in inducing and maintaining a chronic infection state. The results of these three analyses are given below with the highest 10 $\mu^*$ values within each test being highlighted in green and the bottom 10 being highlighted in red. An additional table is provided to show nominal fixed parameters.

\begin{table}[ht!]
\centering
\caption{Fixed parameters for Morris sensitivity analysis.}
\begin{tabular}{|c|c|}
    \hline
     \textbf{Parameter} & \textbf{Nominal Value} \\ \hline
     $N_{\infty}$ & $2 \times 10^7$ \\ \hline
     $k$ & 3 \\ \hline
     $\beta_N$ & $10^{-3}$ \\ \hline
     $d_P$ & 0.99 \\ \hline
     $d_A$ & 0.99 \\ \hline
     $g_N$ & 0.1 \\ \hline
     $\Gamma$ & $5 \times 10^{-4}$ \\ \hline
     $\delta$ & 0.2 \\ \hline
\end{tabular}
\end{table}

\setlength\LTcapwidth{\textwidth}
\begin{longtable}{|l|c|c|c|c|c|c|}
  \caption{Morris sensitivity results.}\label{tab:mu_sigma_values1}\\
  \hline
   & \multicolumn{2}{c|}{Test 1} & \multicolumn{2}{c|}{Test 2} & \multicolumn{2}{c|}{Test 3} \\ 
  \hline
  \textbf{Parameter} & $\mu^*$ & $\sigma$ & $\mu^*$ & $\sigma$ & $\mu^*$ & $\sigma$ \\ 
  \hline
  \endfirsthead

  \multicolumn{7}{@{}l}{\small\itshape Table \ref{tab:mu_sigma_values1} (continued)}\\
  \hline
   & \multicolumn{2}{c|}{Test 1} & \multicolumn{2}{c|}{Test 2} & \multicolumn{2}{c|}{Test 3} \\ 
  \hline
  \textbf{Parameter} & $\mu^*$ & $\sigma$ & $\mu^*$ & $\sigma$ & $\mu^*$ & $\sigma$ \\ 
  \hline
  \endhead

  \hline \multicolumn{7}{r}{\small\itshape Continued on next page}\\
  \endfoot

  \hline
  \endlastfoot

  $C_{QE}$       & \cellcolor[HTML]{cc0000}0.03 & 0.36 & \cellcolor[HTML]{cc0000}0.19 & 2.70  & \cellcolor[HTML]{cc0000}0.07 & 1.29 \\ \hline
  $d_M$         & \cellcolor[HTML]{cc0000}0.11 & 1.05 & \cellcolor[HTML]{cc0000}0.76 & 15.13 & 1.93                        & 22.49 \\ \hline
  $C_{ME}$      & \cellcolor[HTML]{cc0000}0.16 & 2.44 & \cellcolor[HTML]{cc0000}0.03 & 0.29  & \cellcolor[HTML]{cc0000}0.11 & 2.05  \\ \hline
  $C_{UP}$      & \cellcolor[HTML]{cc0000}0.21 & 3.17 & \cellcolor[HTML]{cc0000}0.01 & 0.15  & \cellcolor[HTML]{cc0000}0.05 & 0.61  \\ \hline
  $d_{EN}$      & \cellcolor[HTML]{cc0000}0.27 & 3.24 & \cellcolor[HTML]{cc0000}0.12 & 1.39  & \cellcolor[HTML]{cc0000}1.01 & 19.46 \\ \hline
  $S_{KD}$      & \cellcolor[HTML]{cc0000}0.33 & 4.43 & 1.15                        & 23.09 & 1.96                        & 27.23 \\ \hline
  $C_{UE}$      & \cellcolor[HTML]{cc0000}0.38 & 5.02 & \cellcolor[HTML]{cc0000}0.14 & 1.05  & \cellcolor[HTML]{cc0000}0.37 & 7.25  \\ \hline
  $S_{AS}$      & \cellcolor[HTML]{cc0000}0.51 & 4.07 & 2.25                        & 29.93 & 2.52                        & 33.06 \\ \hline
  $S_{PQ}$      & \cellcolor[HTML]{cc0000}0.52 & 4.86 & \cellcolor[HTML]{cc0000}0.16 & 1.49  & 1.40                        & 25.17 \\ \hline
  $S_{KQ}$      & \cellcolor[HTML]{cc0000}0.54 & 8.43 & 2.08                        & 24.72 & \cellcolor[HTML]{cc0000}0.57 & 4.04  \\ \hline
  $S_{EN}$      & 0.64                        & 7.71 & 2.48                        & 36.91 & \cellcolor[HTML]{cc0000}0.32 & 3.04  \\ \hline
  $R_{KU}$      & 1.14                        &19.96 & 2.39                        & 33.75 & 6.15                        &57.32  \\ \hline
  $S_{AM}$      & 1.16                        &10.83 & 1.47                        & 27.46 & 1.90                        &28.04  \\ \hline
  $d_Q$         & 1.31                        &27.71 & \cellcolor[HTML]{cc0000}0.54 & 5.70  & \cellcolor[HTML]{cc0000}0.19 & 1.53  \\ \hline
  $K_{NM}$      & 1.41                        &13.69 & 1.46                        & 23.30 & 2.74                        &35.43  \\ \hline
  $S_{KMD}$     & 1.72                        &24.80 & 4.10                        & 41.42 & 1.68                        &20.78  \\ \hline
  $S_{AU}$      & 2.08                        &22.49 & 4.62                        & 47.91 & 3.79                        &42.49  \\ \hline
  $I_{\text{crit}}$    & 2.31                        &28.38 & 3.68                        & 40.74 & 3.95                        &41.14  \\ \hline
  $\theta_K$    & 2.60                        &32.29 & 2.17                        & 30.26 & 1.84                        &30.37  \\ \hline
  $K_{NQ}$      & 2.86                        &34.87 & 2.47                        & 36.92 & 1.74                        &19.60  \\ \hline
  $K_{SN}$      & 3.40                        &37.85 & \cellcolor[HTML]{cc0000}0.93 & 18.09 & 2.72                        &34.09  \\ \hline
  $N_{1/2}$     & 3.66                        &41.30 & 1.97                        & 23.02 & \cellcolor[HTML]{cc0000}1.22 &10.32  \\ \hline
  $\Omega$      & 3.68                        &42.94 & 6.60                        & 44.43 & 3.51                        &38.80  \\ \hline
  $d_U$         & 3.79                        &43.94 & 3.74                        & 37.99 & 3.19                        &37.02  \\ \hline
  Infection Size & 3.82                       &39.14 & \cellcolor[HTML]{66ff66}7.06 & 60.34 & 3.87                        &43.43  \\ \hline
  $H_{crit}$    & 4.72                        &46.62 & \cellcolor[HTML]{66ff66}6.24 & 56.23 & \cellcolor[HTML]{66ff66}7.59 &65.35  \\ \hline
  $\tau_Q$      & 4.77                        &49.18 & 2.73                        & 39.11 & \cellcolor[HTML]{cc0000}0.66 & 7.87  \\ \hline
  $S_{PH}$      & 5.19                        &47.38 & \cellcolor[HTML]{66ff66}14.82 & 90.88 & \cellcolor[HTML]{66ff66}9.10 &67.98  \\ \hline
  $\tau_U$      & 5.25                        &47.24 & 3.13                        & 31.89 & 3.47                        &41.65  \\ \hline
  $S_{AH}$      & 6.16                        &53.27 & \cellcolor[HTML]{66ff66}9.92 & 71.86 & \cellcolor[HTML]{66ff66}12.23 &82.50  \\ \hline
  $S_{PS}$      & 6.41                        &54.51 & 4.61                        & 49.10 & 6.46                        &55.07  \\ \hline
  $d_S$         & \cellcolor[HTML]{66ff66}6.57 &59.13 & \cellcolor[HTML]{66ff66}9.04 & 65.63 & 3.92                        &40.55  \\ \hline
  $\theta_N$    & \cellcolor[HTML]{66ff66}6.94 &55.11 & 3.84                        & 37.67 & \cellcolor[HTML]{66ff66}9.74 &72.76  \\ \hline
  $K_{crit}$    & \cellcolor[HTML]{66ff66}7.004 &61.46 & \cellcolor[HTML]{cc0000}0.88 & 15.33 & 2.19                        &30.52  \\ \hline
  $\Psi$        & \cellcolor[HTML]{66ff66}7.33 &60.11 & \cellcolor[HTML]{66ff66}8.16 & 66.37 & \cellcolor[HTML]{66ff66}9.31 &69.97  \\ \hline
  $S_{SCSF}$    & \cellcolor[HTML]{66ff66}9.04 &64.66 & 3.73                        & 39.46 & \cellcolor[HTML]{66ff66}8.88 &66.44  \\ \hline
  $d_{SCSF}$    & \cellcolor[HTML]{66ff66}10.77 &74.58 & \cellcolor[HTML]{66ff66}12.34 & 78.88 & 7.18                        &62.65  \\ \hline
  $C_{hs}$      & \cellcolor[HTML]{66ff66}11.70 &75.92 & 6.00                        & 51.23 & \cellcolor[HTML]{66ff66}8.09 &58.72  \\ \hline
  $K_{NS}$      & \cellcolor[HTML]{66ff66}11.74 &77.00 & \cellcolor[HTML]{66ff66}10.04 & 72.68 & \cellcolor[HTML]{66ff66}9.00 &68.67  \\ \hline
  $\alpha$      & \cellcolor[HTML]{66ff66}12.40 &81.31 & \cellcolor[HTML]{66ff66}7.68 & 61.48 & \cellcolor[HTML]{66ff66}7.73 &59.71  \\ \hline
  $d_H$         & \cellcolor[HTML]{66ff66}12.42 &77.88 & \cellcolor[HTML]{66ff66}7.09 & 62.11 & \cellcolor[HTML]{66ff66}7.58 &63.83  \\ \hline
\end{longtable}

 What immediately stands out is the relationship between each parameter's $\mu^*$ and $\sigma$, namely that the latter is often an entire order-of-magnitude larger than the former. In the context of the Morris Sensitivity test, this most clearly suggests a highly non-linear model where interactions between parameters and variables dominate any potential first-order effect. Given the structure of the model and the biological context it attempts to map onto, this is not an unexpected result.

 Among the parameters screened for being least influential on chronic infection duration, those that consistently show up are those parameters involved in or related to $EN$, implying that the system's ability to sustain robust mature pro-inflammatory cells is likely not a strong factor in the development and maintenance of a chronic infection. These parameters include $C_{iE}$ terms; which control how many $EN$ are consumed \textit{per} cell \textit{i}; $d_{EN}$, which is a standard decay term for $EN$; and $S_{EN}$, which controls the base up-regulation rate of $EN$ in the system. Other notable mentions include $S_{PQ}$, mature pro-inflammatory cell's ability to increase pro-inflammatory signal concentration; $S_{KQ}$, how much tissue damage is incurred by the system by the presence of mature pro-inflammatory cells; and $d_{Q}$, a standard decay term for the pro-inflammatory cells. 

 On the other hand, those parameters that were consistently flagged as being of high influence typically were related to $H_{ST}$, $MPP$, and $S$. For example, $\Psi$, which was flagged as highly influential in all three of the tests, is involved in controlling the rate of differentiation of $S \rightarrow Q, \: U, \: MDSC$; $\alpha$ controls a similar process in $H_{ST}, \:MPP$, it controls the maximum rate of differentiation $H_{ST} \rightarrow MPP \rightarrow S$; both $S_{PH}$ and $S_{AH}$ were flagged twice, these parameters control the extent to which $MPP$'s generate pro-inflammatory / anti-inflammatory signals respectively; $C_{hs}, \: S_{SCSF}, \:$ and $d_{SCSF}$ are all involved in the upregulation of Stem Cell supporting factors and the rate of consumption of these factors by $H_{ST}$, ultimately controlling the robustness of the short-term HSPC pool size; $H_{crit}$ controls how sensitive short-term HSPCs are to increasing their cell-cycle in response to pro-inflammatory signals.

 These results strongly suggest that the \textbf{key driver of chronic infection is the beginning of the hematopoietic tree - the stem cells and immature progenitor cells - rather than the mature cell compartments that are largely implicated in the acute immune response}.
 
 Using the results of the Morris sensitivity tests, we also ran one Sobol sensitivity analysis. While more computationally expensive, a Sobol test can give information on the interactions between parameters. Because running a Sobol test on the full parameter space of this model would be resource intensive, we instead utilized the results of the Morris tests to inform which parameters to freeze. Specifically, we froze all parameters except any that manage to get flagged as highly influential in at least one test, which ended up giving 14 unique parameters to test: Infection Size, $H_{crit}$, $S_{PH}$, $S_{AH}$, $d_S$, $\theta_N$, $K_{crit}$, $\Psi$, $S_{SCSF}$, $d_{SCSF}$, $C_{hs}$, $K_{NS}$, $\alpha$, and $d_H$. The results of the Sobol analysis are presented below.

\begin{table}[ht!]
    \centering
    \caption{First-order and Total-order sensitivity indices computed using Sobol analysis}
    \begin{tabular}{|l|c|c|}
    \hline
    \textbf{Parameter} & \textbf{First-order Index} & \textbf{Total-order Index} \\ \hline
    $H_{crit}$ & -0.0018 & \cellcolor[HTML]{ff8c1a} 0.52 \\ \hline
    $K_{crit}$ & 0.0001 & \cellcolor[HTML]{b30000} 0.0032 \\ \hline
    $S_{AH}$ & 0.0029 & \cellcolor[HTML]{70db70} 0.88 \\ \hline
    $S_{PH}$ & -0.0006 & \cellcolor[HTML]{33cc33} 0.97 \\ \hline
    $S_{SCSF}$ & -0.05 & \cellcolor[HTML]{ffdb4d} 0.72 \\ \hline
    $\alpha$ & -0.05 & \cellcolor[HTML]{ffdb4d} 0.77 \\ \hline
    $d_H$ & -0.05 & \cellcolor[HTML]{33cc33} 0.91 \\ \hline
    $d_{SCSF}$ & -0.03 & \cellcolor[HTML]{33cc33} 0.98 \\ \hline
    $d_S$ & -0.0005 & \cellcolor[HTML]{70db70} 0.85 \\ \hline
    $C_{hs}$ & -0.05 & 0.\cellcolor[HTML]{33cc33} 96 \\ \hline
    $K_{NS}$ & -0.05 & \cellcolor[HTML]{ffdb4d} 0.73 \\ \hline
    $\Psi$ & 0.001 & \cellcolor[HTML]{ff8c1a} 0.57 \\ \hline
    $\theta_N$ & 0.001 & \cellcolor[HTML]{ff8c1a} 0.51 \\ \hline
    Infection Size & -0.002 & \cellcolor[HTML]{ff704d} 0.34 \\ \hline
    \end{tabular}
\end{table}

 The first-order indices being very small agree with the Morris results - namely that the first-order (i.e. individual) effects of parameters are almost negligible compared to the effects of variables interacting with each other. Indeed, both the Morris and Sobol sensitivity analyses make it clear that it is difficult to point to one clear mechanism and claim it to be the sole driver of chronic infection precisely because of the interconnectedness between system variables. Regardless, there appears to be certain parameters and mechanisms that hold more influence than others. In our Sobol test, the parameter $K_{crit}$, which represents the system's robustness against tissue damage, scored a total-order index of $0.0032$ indicating a very weak influence over the time duration of chronic infection, even when accounting for its higher-order interactions with the other 13 parameters tested in our Sobol analysis. The results of the Morris tests and the Sobol test together indicate that $K_{crit}$ is likely not a strong factor on its own in determining whether a chronic low-grade infection persists. All other parameters tested scored $\geq 0.3$, indicating at least moderate influence. The parameters that stood out were 
\begin{itemize}
    \item $S_{AH}$: The rate at which $MPP$s secrete anti-inflammatory signals
    \item $S_{PH}$: The rate at which $MPP$s secrete pro-inflammatory signals
    \item $d_H$: The rate at which $MPP$s naturally decay (owing to their increased cell cycling speed), a higher value would result in faster loss of $MPP$ and consequently less of them
    \item $d_{SCSF}$: The decay rate of Stem Cell Supporting Factors, a higher value would ultimately result in less $SCSF$s which would result in less $H_{ST}$
    \item $d_S$: The decay rate of immature white blood cells, the intermediate stage between stem cells and fully mature white blood cells
    \item $C_{hs}$: The consumption rate of $SCSF$s by $H_{ST}$, a higher value would require more $SCSF$s to support the same number of $H_{ST}$ hence there is an inverse relationship between this parameter and the number of short-term stem cells the system can support
\end{itemize}
 Aside from these top-scoring parameters, $H_{crit}$, $S_{SCSF}$, $\alpha$, $K_{NS}$, $\Psi$, and $\theta_N$ all scored $\geq 0.5$. The parameters $S_{SCSF}$, $d_{SCSF}$, and $C_{hs}$ all directly control the size of the $H_{ST}$ population, and a high score for all three of these indicates that the robustness of the stem cell pool is an important factor in chronic infection. Furthermore, $H_{crit}$ determines how sensitive $H_{ST}$ are to pro-inflammatory signals (i.e. how quickly they ramp up their self-renewal cell cycling in response to pro-inflammatory signals), $\alpha$ determines the maximum rate of differentiation that can be undertaken by $H_{ST}$ and $MPP$ cells. Together with the high scores of $S_{AH}$ and $S_{PH}$, these results suggest that the stem cells' ability to react to inflammatory signals and manipulate their environment through these signals is a key factor in chronic infection worth further study.

 Of the 91 second-order interactions tested by our Sobol analysis, none managed to reach a score much higher than 0.1. Some notable interactions from among those tested include
\begin{itemize}
    \item $S_{SCSF} \leftrightarrow C_{hs}$: 0.098
    \item $C_{hs} \leftrightarrow K_{NS}$: 0.101
    \item $d_{SCSF} \leftrightarrow C_{hs}$: 0.085
    \item $d_H \leftrightarrow d_{SCSF}$: 0.092
\end{itemize}
Altogether, the Morris and Sobol sensitivity analyses suggest that \textbf{the stem cell compartments are a major driving force in the occurrence of chronic infection}.

\section{Discussion}\label{sec:discussion}
 In this paper, we presented a 12-variable mathematical model of hematopoiesis in a healthy steady state and during an inflammatory response to infection. To the best of our knowledge, this is the first mathematical model that attempts to bridge the gap between host response and the hematopoietic dynamics. Where most models assume some steady background supply of cells, we instead directly model this rate as a function of inflammation and the stem cells, giving rise to interesting cell population dynamics. We achieve this by incorporating distinct compartments of cells, each representing a different stage of maturation/differentiation, and incorporating a mechanism by which the interaction of these cells with pro-inflammatory signals induces further differentiation. We were able to produce a range of clinically relevant behaviors and examine how stem cell dynamics can influence the development of a chronic infection state whereby a low-grade persistent infection presents itself and fails to resolve for extended periods of time, and, in some cases, even gives rise to a new stable fixed point of the system in between pathogen clearance ($N=0$) and pathogens growing to carrying capacity ($N=N_{\infty}$).

 An interesting behavior emerges from the model dynamics, namely that the pro-inflammatory response is inherently self-limiting: after a certain threshold the pro-inflammatory force begins to down-regulate itself (see Fig \ref{fig:aseptic} for an example). This occurs primarily because of the positive relationship between pro-inflammation and tissue damage and the negative relationship between tissue damage and cell populations. The pro-inflammatory response induces a large influx of mature pro-inflammatory white blood cells. These blood cells are known to induce damage to healthy living tissue and pathogens alike, which in our model is tuned by the parameters $S_{KQ}$ and $K_{NQ}$, respectively. As tissue damage increases, the ability of the system to maintain robust healthy cell populations decreases, which results in less mature pro-inflammatory white blood cells which ultimately weakens the pro-inflammatory response. This is important because it suggests that an excessive pro-inflammatory response alone is unlikely to lead to the development of chronic inflammation and instead, it is a combination of this with other factors, such as a reduced capacity for healing tissue damage, that lead to chronic inflammation.  

 %The results of our sensitivity analysis contradict literature on the role MDSCs may play in the development and/or maintenance of chronic infection (e.g. see \cite{sanchez2021myeloid}). Our results suggest that the mechanisms by which MDSCs exert their immunosuppressive influence over the hematopoietic response (by controlling the extent of the pro-inflammatory white blood cell response) holds little influence over the duration of chronic infection relative to other mechanisms, particularly those of the stem cell and multipotent WBC compartments. This may simply be an artifact of our model's structure or it may be indicative of real world dynamics, but regardless we believe the role stem cells and MDSCs play in the development of chronic infection to be worth further study.

 It is important to recognize the limitations of the model presented here. Our model was specifically built to study hematopoiesis, the process by which white blood cells are generated from stem cells to meet the body's demands. To reduce computational complexity, we did not explicitly model spatial dynamics, instead opting to use crowding effects where necessary to emulate this as best we could. Many of the mechanisms we do not include in this model, such as cell migration to sites of infection through the action of chemokines, are related to this lack of a spatial dimension. Furthermore, we utilize certain simplifications to keep the model tractable, such as the categorization of cytokines into a binary pro-inflammatory or anti-inflammatory class, which removes much of the nuance of each particular cytokine (e.g. IL-6 can have either pro- or anti-inflammatory effects depending on the specific context of its activation), and the grouping of cells by function (such as $Q$ and $U$ being the mature pro-inflammatory and anti-inflammatory WBC pools respectively) as opposed to making a distinct class for each type of cell (e.g. a distinct class for neutrophils, T-cells, monocytes, etc.). These limitations may explain why our results contradict currently accepted roles of MDSC prevalence in chronic infection.

 As part of these limitations, we observed that our model does not recapitulate the chronically elevated MDSC population that is associated with particularly severe cases of sepsis under nominal parameter ranges \cite{mathias2017human} \cite{horiguchi2018innate}. This suggests that \textbf{dysfunctional hematopoiesis alone cannot fully explain the persistent MDSC presence observed in CCI} and more work must be done to understand the mechanisms guiding the behavior of MDSCs. The inability of our model to produce this behavior can explain why parameters associated with the MDSC group did not flag as influential in our sensitivity results.

\bibliographystyle{plain}
\bibliography{sources}  % how overleaf expects it

\appendix 
\section{Model Derivation}\label{si:model}

\begin{figure}[ht!]
\includegraphics[width=\textwidth]{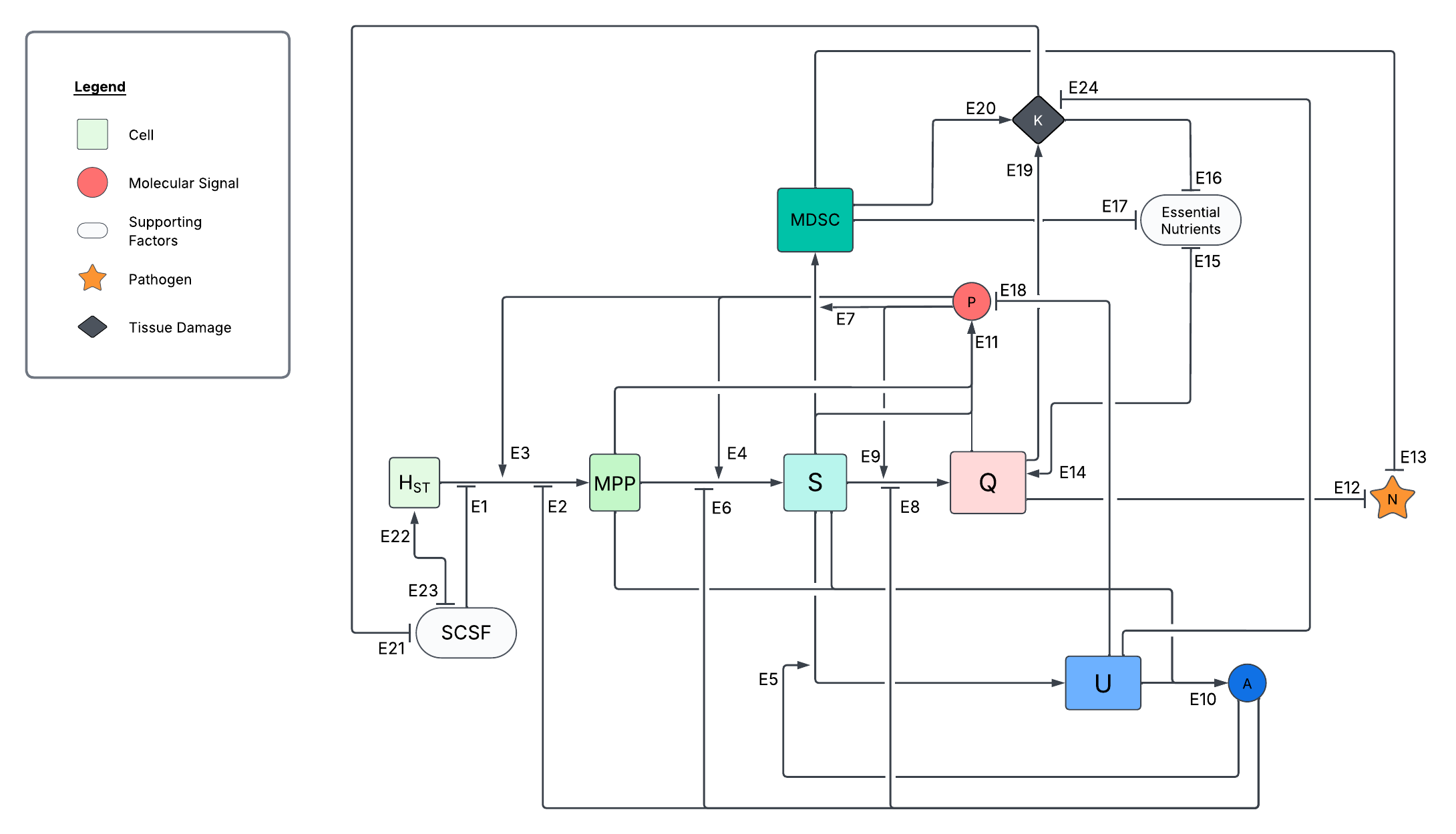}
  \caption{12-variable system figure.}\label{fig:node2}
\end{figure}
A brief explanation of each edge is given below:
\begin{itemize}
    \item \underline{E1:} Adhesion molecules maintain stem cell quiescence and retention in the BM Niche
    \item \underline{E2:} Anti-inflammatory signals maintain stem cell quiescence
    \item \underline{E3:} Pro-inflammatory signals interact with stem cells, causing a transition from quiescence to proliferation
    \item \underline{E4:} Pro-inflammatory signals induce differentiation of stem cells into progenitor cells
    \item \underline{E5:} Anti-inflammatory signals promote the expansion of the anti-inflammatory phenotype of mature white blood cells
    \item \underline{E6:} Anti-inflammatory signals inhibits progenitor cell differentiation / maturation
    \item \underline{E7:} Pro-inflammatory signals induce the expansion of the MDSC population
    \item \underline{E8:} Anti-inflammatory signals inhibits immature hematopoietic cell differentiation / maturation
    \item \underline{E9:} Pro-inflammatory signals enhance the rate of differentiation / maturation of immature hematopoietic cells
    \item \underline{E10:} Mature immuno-suppressive phenotype cells secrete anti-inflammatory signals at an enhanced rate
    \item \underline{E11:} Mature pro-inflammatory phenotype cells secrete pro-inflammatory signals at an enhanced rate
    \item \underline{E12:} Mature pro-inflammatory phenotype cells remove pathogens at an enhanced rate
    \item \underline{E13:} MDSCs remove pathogens, though not as efficiently as mature pro-inflammatory cells
    \item \underline{E14:} Essential nutrients are required to sustain a robust mature pro-inflammatory phenotype cell population
    \item \underline{E15:} Essential nutrients are consumed by mature pro-inflammatory phenotype cells
    \item \underline{E16:} Tissue damage affects the system's ability to up-regulate essential nutrients
    \item \underline{E17:} MDSCs consume essential nutrients as part of their immuno-suppressive abilities
    \item \underline{E18:} Mature immuno-suppressive cells can consume / negate the effects of pro-inflammatory signals as part of their immuno-suppressive abilities
    \item \underline{E19:} Mature pro-inflammatory phenotype cells can cause damage to surrounding tissue at enhanced rates, directly and indirectly
    \item \underline{E20:} MDSCs can indirectly cause damage to surrounding tissue
    \item \underline{E21:} Tissue damage affects the BM Niche's ability to maintain ingredients needed to support stem cell population
    \item \underline{E22:} Stem cell supporting factors are necessary for maintaining stem cell population in BM Niche
    \item \underline{E23:} Stem cells consume stem cell supporting factors
    \item \underline{E24:} Mature immuno-suppressive cells can heal tissue damage and restore system health
\end{itemize}

\begin{itemize}
    \item {$H_{ST}$}: Short-term HSPCs, stem cells that have exited quiescence and have enhanced cell-cycling times and limited self renewal.
    \item {$MPP$}: HSPCs that have exited quiescence and differentiated into \textit{multipotent progenitor} cells (or MPP for short); these cells retain a limited capacity for self-renewal, have an increased self-cycling rate, and are much more reactive to signals from the local inflammatory environment.
    \item {$S$}: Immature WBCs; these white blood cells have lost their self-renewal capacity and have a reduced capacity for cytokine signaling and pathogen removal compared to their mature pro-inflammatory counterparts.
    \item {$Q$ and $U$}: The last stage of maturation; for our reduced hematopoiesis model we simplify WBCs into either a pro-inflammatory phenotype $Q$ or anti-inflammatory phenotype $U$.
    \item $K$ and $EN$: $K$ represents \textit{Damage Associated Molecular Patterns}, hereby abbreviated with its commonly used name DAMPs, are a class of endogenous molecules that are released by damaged or dying cells \cite{roh2018damage,ma2024damps}. $EN$ is short for \textit{Essential Nutrients}.
    \item $MDSC$: $MDSC$ stands for \textit{myeloid-derived suppressor cell}, a name used to refer to a heterogeneous group of immature myeloid cells with potent immuno-suppressive abilities \cite{gabrilovich2009myeloid,veglia2021myeloid}.
\end{itemize}

\section{Parameters}\label{si:parameters}

\begin{table}[ht!]
  \centering
  \caption{Parameters, descriptions, units, and nominal values}
  \label{tab:parameters}
  \begin{tabular}{|c|p{0.4\textwidth}|c|c|}
    \hline
    \textbf{Parameter} & \textbf{Description} & \textbf{Units} & \textbf{Nominal Value} \\
    \hline
    $C_{hs}$   & Rate of consumption of $SCSF$s per HSPC cell                                                 & $SCSF\,H_{ST}^{-1}$       & $3$    \\ \hline
    $d_H$      & Decay rate of mobilized HSPCs                                                                 & $MPP\,\mathrm{hr}^{-1}$   & $0.05$ \\ \hline
    $\theta_N$ & Pathogen concentration at half‐maximal pro‐inflammatory signal amplification                    & $N$                       & $100\,000$ \\ \hline
    $\theta_K$ & Tissue‐damage concentration at half‐maximal pro‐/anti‐inflammatory signal amplification         & $K$                       & $50\,000$  \\ \hline
    $\tau_Q$   & Sensitivity coefficient of immature WBCs to pro‐inflammatory signals                     & Non‐dimensional           & $1$    \\ \hline
    $\tau_U$   & Sensitivity coefficient of immature WBCs to anti‐inflammatory signals                    & Non‐dimensional           & $1$    \\ \hline
    $d_S$      & Decay rate of immature WBCs                                                             & $\mathrm{hr}^{-1}$        & $0.15$ \\ \hline
    $d_Q$      & Decay rate of pro‐inflammatory WBCs                                                     & $\mathrm{hr}^{-1}$        & $0.95$ \\ \hline
    $d_U$      & Decay rate of anti‐inflammatory WBCs                                                    & $\mathrm{hr}^{-1}$        & $0.2$  \\ \hline
    $d_P$      & Decay rate of pro‐inflammatory signals                                                        & $\mathrm{hr}^{-1}$        & $0.99$\,\cite{whiteside1994cytokines,liu2021cytokines} \\ \hline
    $d_A$      & Decay rate of anti‐inflammatory signals                                                       & $\mathrm{hr}^{-1}$        & $0.99$\,\cite{whiteside1994cytokines,liu2021cytokines} \\ \hline
    $S_{PH}$   & Secretion rate of pro‐inflammatory signals by mobilized HSPCs                                 & $P/(MPP\cdot hr)$         & $2$    \\ \hline
    $S_{PS}$   & Secretion rate of pro‐inflammatory signals by immature WBCs                             & $P/(S\cdot hr)$           & $5$    \\ \hline
    $S_{PQ}$   & Secretion rate of pro‐inflammatory signals by pro‐inflammatory WBCs                     & $P/(Q\cdot hr)$           & $10$   \\
    $C_{UP}$   & Consumption rate of pro‐inflammatory signals by anti-inflammatory WBCs                  & $\tfrac{P}{U\cdot hr}$         & $2$    \\ \hline
    \hline
  \end{tabular}
  
\end{table}

\begin{table}[ht!]
  \centering
  \caption{Nominal parameter values}
  \label{tab:nominal_params}
  \begin{tabular}{|c|p{0.4\textwidth}|c|c|}
    \hline
    \textbf{Parameter} & \textbf{Description}  & \textbf{Units} & \textbf{Nominal Value} \\
    \hline
    $S_{AU}$    & Secretion rate of anti‐inflammatory signals by anti‐inflammatory WBCs                  & $\tfrac{A}{U\cdot hr}$ & $15$           \\
    \hline
    $S_{AS}$    & Secretion rate of anti‐inflammatory signals by immature WBCs                           & $\tfrac{A}{S\cdot hr}$ & $3$            \\
    \hline
    $S_{AH}$    & Secretion rate of anti‐inflammatory signals by mobilized HSPCs                               & $\tfrac{A}{MPP\cdot hr}$ & $1$           \\
    \hline
    $I_{\rm crit}$ & Inflammation metric value at which half of mobilized HSPCs undergo symmetric differentiation & $\tfrac{P}{P+A}$        & $0.4$          \\
    \hline
    $\Psi$      & Rate at which signals spread throughout the immature WBC population                    & \emph{Non‐dimensional} & $0.4$          \\
    \hline
    $\alpha$    & Sensitivity coefficient of HSPCs to pro‐inflammatory signals                                & \emph{Non‐dimensional} & $0.3$          \\
    \hline
  \end{tabular}
\end{table}

\begin{table}[ht]
  \centering
  \caption{Additional model parameters and their nominal values}
  \label{tab:additional_parameters}
  \begin{tabular}{|c|p{0.4\textwidth}|c|c|}
    \hline
    \textbf{Parameter} & \textbf{Description} & \textbf{Units} & \textbf{Nominal Value} \\
    \hline
    $k$  & Hill coefficient in the amplification functions  & \emph{Non‐dimensional} & $3$               \\ \hline
    $g_N$  & Proliferation rate of pathogens  & $\mathrm{hr}^{-1}$     & $0.2$             \\ \hline
    $N_{\infty}$  & Carrying capacity of pathogens & $N$  & $2\times10^{7}$   \\ \hline
    $N_{1/2}$     & Pathogen concentration at which removal rate by WBCs is half its maximum   & $N$  & $2\,500$ \\ \hline
    $K_{SN}$      & Removal rate of immature WBCs per pathogen & $\tfrac{S}{N\cdot hr}$ & $3$ \\ \hline
    $K_{NQ}$   & Removal rate of pathogens by pro‐inflammatory WBCs  & $\tfrac{N}{Q\cdot hr}$ & $10$ \\ \hline
    $K_{NS}$      & Removal rate of pathogens by immature WBCs  & $\tfrac{N}{Q\cdot hr}$ & $10$ \\ \hline
  \end{tabular}
\end{table}

\begin{table}[ht]
  \centering
  \caption{Model parameters with nominal values and ranges}
  \label{tab:additional_model_parameters}
  \begin{tabular}{|c|p{0.4\textwidth}|c|c|}
    \hline
    \textbf{Parameter} & \textbf{Description} & \textbf{Units} & \textbf{Nominal Value / Range} \\
    \hline
    $S_{SCSF}$ & Maximum rate of up‐regulation by background tissue of essential ingredients for supporting HSPC populations
                & $\tfrac{SCSF}{K\cdot hr^{-1}}$ & $10000$ \\
    \hline
    $\Gamma$ & Minimum proliferation rate of HSPCs
              & $\mathrm{hr}^{-1}$ & $5\times10^{-4}$\,\cite{scala2019vivo,osorio2018somatic,lee2018population,boyle2023predicting} \\
    \hline
    $\Delta$ & Maximum proliferation rate of HSPCs & $\mathrm{hr}^{-1}$ & $0.2$\,\cite{scala2019vivo,osorio2018somatic,lee2018population,boyle2023predicting}\\
    \hline
    $S_{KD}$ & Rate of tissue damage increase as a direct result of infection
              & $\tfrac{K}{N\cdot hr}$ & $3$ \\
    \hline
    $R_{KU}$ & Rate of healing of tissue damage by anti‐inflammatory WBCs
              & $\tfrac{K}{U\cdot hr}$ & $10$ \\
    \hline
    $d_{SCSF}$ & Decay rate of SCSFs
               & $\mathrm{hr}^{-1}$ & $0.3$ \\
    \hline
    $K_{crit}$ & Tissue‐damage level at which background up‐regulation of cellular nutrients is half‐maximal
                & $K$ & $150\,000$ \\
    \hline
    $S_{EN}$ & Maximum rate of up‐regulation by background tissue of essential ingredients for supporting pro‐inflammatory WBC populations
              & $\tfrac{EN}{K\cdot hr^{-1}}$ & $500$ \\
    \hline
    $S_{KQ}$ & Rate of tissue damage increase as a direct result of pro‐inflammatory WBC activity
              & $\tfrac{K}{Q\cdot hr}$ & $8$ \\
    \hline
    $S_{KQ}$ & Rate of tissue damage increase as a direct result of pro‐inflammatory WBC activity
              & $\tfrac{K}{Q\cdot hr}$ & $8$ \\
    \hline
    $C_{QE}$ & Consumption rate of cellular ingredients by pro‐inflammatory WBCs
              & $\tfrac{EN}{Q\cdot hr}$ & $3$ \\
    \hline
    $C_{UE}$ & Consumption rate of cellular ingredients by anti‐inflammatory WBCs
              & $\tfrac{EN}{U\cdot hr}$ & $3$ \\
    \hline
    $d_{EN}$ & Decay rate of molecular factors
              & $\mathrm{hr}^{-1}$ & $0.3$ \\
    \hline
    $H_{crit}$ & Value of inflammation ($I_H$) at which HSPC proliferation rate reaches half maximum velocity 
              & \textit{Non-dimensional} & $0.2$ \\
    \hline
  \end{tabular}
  
\end{table}

\begin{table}[ht]
  \centering
  \caption{MDSC‐related model parameters and their nominal values}
  \label{tab:mdsc_parameters}
  \begin{tabular}{|c|p{0.4\textwidth}|c|c|}
    \hline
    \textbf{Parameter} & \textbf{Description}                                                            & \textbf{Units}               & \textbf{Nominal Value} \\
    \hline
    $S_{AM}$ & Secretion rate of anti‐inflammatory signals per MDSC                                  & $\tfrac{A}{\mathrm{MDSC}\cdot hr}$ & $12$                   \\ \hline
    $d_M$    & Decay rate of MDSCs                                                                  & $\mathrm{hr}^{-1}$             & $0.9$                  \\ \hline
    $S_{KMD}$ & Increase in tissue damage as a direct result of MDSC activity                         & $\tfrac{K}{\mathrm{MDSC}\cdot hr}$ & $2$                    \\ \hline
    $C_{ME}$ & Consumption rate of molecular factors by MDSCs                                       & $\tfrac{EN}{\mathrm{MDSC}\cdot hr}$ & $5$                    \\ \hline
    $\Omega$ & Proportion of interactions $(S+P)$ that result in $S\rightarrow Q$                    & \emph{Non‐dimensional}         & $0.7$                  \\ \hline
    $K_{NM}$ & Removal rate of pathogens by MDSCs                                                   & $\tfrac{N}{\mathrm{MDSC}\cdot hr}$ & $3$                    \\ \hline
  \end{tabular}
  
\end{table}

\end{document}